\documentclass[12pt,amsmath,amssymb,aps,nofootinbib]{revtex4-1}
\usepackage{lipsum}
\usepackage{graphicx,epsfig,dcolumn,bm,epic,eepic,float}
\usepackage[T1]{fontenc}
\usepackage[utf8]{inputenc}
\usepackage{mathtools} 
\usepackage{pifont}
\usepackage{amsmath}
\usepackage{latexsym}
\usepackage{color}
\usepackage{cancel}
\usepackage{scrextend}
\usepackage{amsmath}
\usepackage{amsthm}
\usepackage{eucal}
\usepackage{amssymb}
\usepackage{mathrsfs}
\usepackage[bottom]{footmisc}
\usepackage{makeidx,shortvrb,latexsym}
\usepackage{epstopdf}
\usepackage{mathtools}
\usepackage{ragged2e}
\usepackage[T1]{fontenc}
\usepackage{lipsum}
\usepackage{multirow}
\usepackage{array}
\usepackage{hyperref}
\newcounter{rowno}
\makeatletter
\newcommand*{\rom}[1]{\expandafter\@slowromancap\romannumeral #1@}
\newcommand {\be}{\begin{equation}}
\newcommand {\ee}{\end{equation}}
\newcommand {\ba}{\begin{eqnarray}}
\newcommand {\ea}{\end{eqnarray}}
\newcommand {\bea}{\begin{eqnarray}}
\newcommand {\eea}{\end{eqnarray}}

\pdfoutput=1

\numberwithin{equation}{section}
\begin{document}

\begin{flushright}
MAN/HEP/2019/009\\
December 2019
\end{flushright}
\vspace*{0.7cm}

\title{{\LARGE Classifying Accidental Symmetries} \\[3mm]
{\LARGE in Multi-Higgs Doublet Models}\\[1mm]
${}$}

\author{\large Neda Darvishi$\,$} \email{neda.darvishi@manchester.ac.uk}
\author{\large Apostolos Pilaftsis$\,$}\email{apostolos.pilaftsis@manchester.ac.uk}

\affiliation{~~~~~~~~~~~~~~~~~~~~~~~~~~~~~~~~~~~~~~~~~~~~~~~~~~~~~~~~~~~~${}$\vspace{-3mm}\\
Consortium for Fundamental Physics, School of Physics and
 Astronomy,\\University of Manchester, Manchester M13 9PL, United
 Kingdom} 
\begin{abstract}
${}$

\centerline{\bf ABSTRACT} \medskip
\noindent
The potential of $n$-Higgs Doublet Models ($n$HDMs) contains a large
number${}$ of SU(2)$_L$-preserving accidental symmetries as subgroups of the symplectic group~Sp(2$n)$. To classify these, we introduce prime
invariants and irreducible representations in bilinear field space
that enable us to explicitly construct accidentally symmetric $n$HDM
potentials. We showcase the classifications of symmetries and present
the relationship among the theoretical parameters of the scalar
potential for the following: (i) the Two Higgs Doublet Model (2HDM), and (ii) the Three Higgs Doublet Model (3HDM). We recover the maximum number of $13$ accidental symmetries for the 2HDM potential, and {\em for the first time}, we present the complete list of $40$ accidental
symmetries for the 3HDM potential.

\end{abstract}
\maketitle
\section{Introduction}\label{sec:intro}

The discovery of the Higgs particle at the CERN Large Hadron Collider
(LHC) \cite{Aad:2012tfa,Chatrchyan:2012xdj}, which was previously
predicted by the Standard Model (SM) of Particle Physics
\cite{Glashow:1961tr,Weinberg:1967tq,Salam1968,Englert:1964et,Higgs:1964pj},
generated renewed interest in Beyond the SM (BSM) Higgs Physics. This
is corroborated by the fact that the SM fails to address several key
questions, such as the origin of the observed matter-antimatter
asymmetry and the dark matter relic abundance in the Universe.

There is a plethora of well-motivated New Physics models with
additional Higgs scalars that have been introduced to solve these
problems~\cite{Lee1,Ginzburg:1999fb,Branco1,Delgado:2013zfa,Bonilla:2014xba,
  Darvishi:2016tni}. To distinguish such models, one usually employs
symmetry transformations that leave the particular model
invariant. These symmetries impose constraints over the theoretical
parameters of the models and thus enhance their predictability to be
probed in future experiments.

In this paper we construct potentials of multi-Higgs Doublet Models
($n$HDMs) with $n$ scalar-doublets based on SU(2)$_L$-preserving
accidental symmetries. These symmetries can be realized in two sets:
(i) as continuous symmetries and (ii) as discrete symmetries (Abelian and
non-Abelian symmetry groups). To find continuous symmetries, we
present an algorithmic method that provides the full list of proper,
improper and semi-simple subgroups for any given integer $n$. We also
include all known discrete symmetries in $n$HDM
potentials~\cite{Derman:1977wq,
  Pakvasa:1977in,Pakvasa:1978tx,Yamanaka:1981pa,Brown:1984mq,Segre:1978ji,Ishimori:2012zz,Ivanov:2012fp,Keus:2013hya,Ivanov:2014doa,deMedeirosVarzielas:2019rrp}. Previous
analyses led to a maximum of 13 accidental symmetries for the Two
Higgs Doublet Model (2HDM) potential, where the maximal symmetry group
is
$G^{\bm{\Phi}}_ {\text{2HDM}}=\text{Sp(4)}/ \text{Z}_2 \otimes
\text{SU(2)}_L$~\cite{Pilaftsis:2011ed}.
Here, we extend this theoretical framework to $n$HDM potentials based
on the maximal symmetric group
Sp($2n)/Z_2
\otimes\text{SU(2)}_L$~\cite{Pilaftsis:2016erj,Darvishi:2019ltl}.
The Maximally Symmetric $n$HDM (MS-$n$HDM) can provide natural SM
alignment that exhibits quartic coupling unification up to the
Planck scale~\cite{Dev:2014yca,Pilaftsis:2016erj,Darvishi:2019ltl,Darvishi:2020teg}.

Identifying the symplectic group Sp($2n)$ as the maximal symmetry
group allows us to classify all SU(2)$_L$-preserving accidental
symmetries in $n$HDM potentials.  We introduce {\em{prime invariants}}
to construct accidentally symmetric potentials in terms of fundamental
building blocks that respect the symmetries. In the same context, we
use irreducible representations to derive all potentials that are
invariant under non-Abelian discrete symmetries.

The layout of this paper is as follows. In Section \ref{sec:s1}, we
define the $n$HDMs in the bilinear scalar field formalism. Given that
Sp(2$n)$ is the maximal symmetry of the $n$HDM potential, we adopt
the bi-adjoint representation of this group which becomes relevant to this bilinear formalism.  In Section
\ref{sec:cs}, we start with classifying continuous symmetries for
$n$HDM potentials. As illustrative examples, we classify all
accidentally symmetric potentials for the 2HDM and the Three Higgs
Doublet Model (3HDM). Then, we introduce prime invariants to build
potentials that are invariant under SU(2)$_L$-preserving continuous
symmetries. In Section~\ref{sec:ds}, we discuss possible and known
discrete symmetries for $n$HDM potentials and recover the discrete
symmetries for the 2HDM and the 3HDM cases
\cite{Ivanov:2012fp,Ivanov:2014doa,deMedeirosVarzielas:2019rrp}. This
section also includes our approach to building 3HDM potentials with
the help of irreducible representations of discrete symmetries. Having
discussed the classifications of symmetries, we provide the list of
all SU(2)$_L$-preserving accidental symmetries for the 2HDM and the
3HDM potentials including the relationships among the theoretical
parameters of the scalar potentials. Section \ref{sec:con} contains
our conclusions. Finally, technical details are delegated to
Appendices \ref{V2-3}, \ref{sp-g}, \ref{C}, and \ref{D}.

\section{Bilinear Formalism and the Maximal Symmetry Sp($2n$)}
\label{sec:s1}

The $n$HDMs contain $n$ scalar doublet fields,
$ \phi_{ i} \,\, (i=1,2,\dots,n)$, which all share the same
U(1)$_Y$-hypercharge quantum number, i.e. $Y_{\phi_{ i} }={1/2}$.  The
most general form of the $n$HDM potential may conventionally be
expressed as follows \cite{Botella:1994cs}:
\begin{equation}
V_n=\ \displaystyle\sum_{i,j=1}^{n}\, m_{ij}^2\, ( \phi_i^{\dagger} \phi_j)+ \ \displaystyle\sum_{i,j\, k,l=1}^{n}\, 
\lambda_{ijkl}\, ( \phi_i^{\dagger} \phi_j)( \phi_k^{\dagger} \phi_l),
\end{equation}
with $\lambda_{ijkl}=\lambda_{klij}$.  In general, the
SU(2)$_L\times$U(1)$_Y$ invariant $n$HDM potential contains $n^2$
physical mass terms along with $n^2(n^2+1)/2$ physical quartic
couplings.

An equivalent way to write the $n$HDM potential is based on the
so-called bilinear field formalism
\cite{Velhinho:1994np,Maniatis:2006fs,Nishi:2006tg,Ivanov:2006yq,Battye:2011jj}.  To
this end, we first define a $4n$-dimensional ($4n$-D) complex
${\bm{\Phi}}_n$-multiplet as
\begin{equation}
\small
{\bm{\Phi}}_2 =\begin{pmatrix}
\phi_1 \\
\phi_2 \\
i \sigma^2 \phi_1^{*} \\
i \sigma^2 \phi_2^{*}
\end{pmatrix}, \quad
{\bm{\Phi}}_3 =\begin{pmatrix}
\phi_1 \\
\phi_2 \\
\phi_3 \\
i \sigma^2 \phi_1^{*} \\
i \sigma^2 \phi_2^{*} \\
i \sigma^2 \phi_3^{*} 
\end{pmatrix},\quad \dots,\quad
{\bm{\Phi}}_n =\begin{pmatrix}
\phi_1 \\
\phi_2 \\
\phi_3 \\
\cdot \\
\cdot \\
\cdot \\
i \sigma^2 \phi_1^{*} \\
i \sigma^2 \phi_2^{*} \\
i \sigma^2 \phi_3^{*} \\
\cdot \\
\cdot \\
\cdot 
\end{pmatrix},
\end{equation}
where $\sigma^{1,2,3}$ are the Pauli matrices and
$i \sigma^2 \phi_i^{*}$ (with $i=1,2,\dots,n$) are the U(1)$_Y$
hypercharge-conjugate of $\phi_i$.  Observe that the
${\bm{\Phi}}_n$-multiplet transforms covariantly under an SU(2$)_L$
gauge transformation as,
\begin{equation}
{\bm{\Phi}}_n\to \text{U}_L {\bm \Phi}_n, \quad \text{U}_L\in \text{SU(2)}_L.
\label{su(2)}
\end{equation}
Note that U$_L$ is the 2$\times$2 unitary matrix that may also be
represented as $\sigma^0 \otimes {\bf{1}}_n \otimes \text{U}_L$ in the
${\bm{\Phi}}_n$-space.  Additionally, the ${\bm{\Phi}}_n$-multiplet
satisfies the Majorana-type property \cite{Battye:2011jj},
\begin{equation}
{\bm{\Phi}}_n=C {{\bm \Phi}_n^*}, 
\end{equation}
where $C=\sigma^2 \otimes {\bf{1}}_n \otimes \sigma^2$
($C=C^{-1}=C^*$) is the charge conjugation operator and ${\bf{1}}_n$
is the $n\times n$ identity matrix.

With the help of the ${\bm{\Phi}}_n$-multiplet, we may now define the
bilinear field
vector~\cite{Ivanov:2006yq,Pilaftsis:2011ed,Battye:2011jj},
\begin{equation}
R_{n}^{A} \equiv {\bm{\Phi}}_n^{\dagger} {\Sigma}^{A}_n {\bm{\Phi}}_n,
\end{equation}
with ${A}=0,1,2,\dots,n(2n-1)-1$. Notice that $n(2n-1)$-vector
$R_{n}^{A}$ is invariant under SU(2)$_L$ transformations thanks to
(\ref{su(2)}).

The matrices ${\Sigma}^{A}_n$ have $4n \times 4n$ elements and
can be expressed in terms of double tensor products as,
\begin{gather}
{\Sigma}^{A}_n=\left(\sigma^0 \otimes t_S^a \otimes \sigma^0,
 \quad \sigma^i \otimes t_A^b \otimes \sigma^0 \right),
\end{gather}
where $t_S^a$ and $ t_A^b$ are the symmetric and anti-symmetric
matrices of the SU($n$) symmetry generators, respectively. Specifically,
for the case of the 2HDM, the following 6 matrices may be defined:
\begin{eqnarray}
{\Sigma_2}^{0,1,3}&=&\dfrac{1}{2}\sigma^0 \otimes \sigma^{0,1,3} \otimes \sigma^0, \quad \quad {\Sigma_2}^{2}=\dfrac{1}{2}\sigma^3 \otimes \sigma^{2} \otimes \sigma^0, 
\nonumber \\
{\Sigma_2}^{4}&=&-\dfrac{1}{2}\sigma^2 \otimes \sigma^{2} \otimes \sigma^0, \quad \quad \;
 {\Sigma_2}^{5}=-\dfrac{1}{2}\sigma^1 \otimes \sigma^{2} \otimes \sigma^0.
\end{eqnarray}
Correspondingly, for the 3HDM, we have the following 15 matrices:
\begin{eqnarray}
{\Sigma_3}^{0,1,2,3,7,8}&=&\dfrac{1}{2}\sigma^0 \otimes G^{0,1,4,6,3,8} \otimes \sigma^0,\quad \quad
 {\Sigma_3}^{4,5,6}=\dfrac{1}{2}\sigma^3 \otimes G^{2,5,7} \otimes \sigma^0, \nonumber \\
{\Sigma_3}^{9,10,11}&=&\dfrac{1}{2}\sigma^2 \otimes G^{2,5,7} \otimes \sigma^0, \quad \quad \;\;
{\Sigma_3}^{12,13,14}=\dfrac{1}{2}\sigma^1 \otimes G^{2,5,7} \otimes \sigma^0,
\end{eqnarray}
where $G^i$ are the standard Gell-Mann matrices of SU(3)~\cite{GellMann:1962xb}.
Note that the ${\Sigma}^{A}_n$ matrices satisfy the property,
\bea
C^{-1}{\Sigma}
^{A}_nC=({\Sigma}^{A}_n)^{\mathsf{T}},
\eea
which means that $ {\Sigma}^{A}_n$ are C-even.

Consequently, the vectors $R_{2}^{A}$ and $R_{3}^{A}$ for the 2HDM and
the 3HDM cases are given by
\begin{equation}
\small  
R^{A}_2= \begin{pmatrix}
\phi_1^{\dagger} \phi_1 +\phi_2^{\dagger} \phi_2 \\ 
\phi_1^{\dagger} \phi_2 +\phi_2^{\dagger} \phi_1 \\ 
-i [\phi_1^{\dagger} \phi_2 -\phi_2^{\dagger} \phi_1] \\
\phi_1^{\dagger} \phi_1 -\phi_2^{\dagger} \phi_2 \\
-[\phi_1^{\mathsf{T}} i \sigma^2 \phi_2 - \phi_2^{\dagger} i \sigma^2 \phi_1^* ]\\
i[\phi_1^{\mathsf{T}} i \sigma^2 \phi_2 + \phi_2^{\dagger} i \sigma^2 \phi_1^* ]
\end{pmatrix},\quad
R^{A}_{3}= \begin{pmatrix}
\phi_1^{\dagger} \phi_1 +\phi_2^{\dagger} \phi_2+\phi_3^{\dagger} \phi_3 \\ 
\phi_1^{\dagger} \phi_2 +\phi_2^{\dagger} \phi_1 \\ 
\phi_1^{\dagger} \phi_3 +\phi_3^{\dagger} \phi_1 \\ 
\phi_2^{\dagger} \phi_3+\phi_3^{\dagger} \phi_2\\ 
-i [\phi_1^{\dagger} \phi_2 -\phi_2^{\dagger} \phi_1] \\
-i [\phi_1^{\dagger} \phi_3 -\phi_3^{\dagger} \phi_1] \\
-i [\phi_2^{\dagger} \phi_3 -\phi_3^{\dagger} \phi_2] \\
\phi_1^{\dagger} \phi_1 -\phi_2^{\dagger} \phi_2 \\
{1\over\sqrt{3}}[\phi_1^{\dagger} \phi_1 +\phi_2^{\dagger} \phi_2-2 \phi_3^{\dagger} \phi_3] \\ 
-[\phi_1^{\mathsf{T}} i \sigma^2 \phi_2 - \phi_2^{\dagger} i \sigma^2 \phi_1^*] \\
-[\phi_1^{\mathsf{T}} i \sigma^2 \phi_3 - \phi_3^{\dagger} i \sigma^2 \phi_1^*] \\
-[\phi_2^{\mathsf{T}} i \sigma^2 \phi_3 - \phi_3^{\dagger} i \sigma^2 \phi_2^* ]\\
i[\phi_1^{\mathsf{T}} i \sigma^2 \phi_2 + \phi_2^{\dagger} i \sigma^2 \phi_1^* ]\\
i[\phi_1^{\mathsf{T}} i \sigma^2 \phi_3 + \phi_3^{\dagger} i \sigma^2 \phi_1^* ]\\
i[\phi_2^{\mathsf{T}} i \sigma^2 \phi_3 + \phi_3^{\dagger} i \sigma^2 \phi_2^* ]
\end{pmatrix}.
\end{equation}
With the aid of the $n(2n-1)$-dimensional vectors $R_{n}^{A}$, the
potential $V_n$ for an $n$HDM can be written in the quadratic
form as
\begin{equation}
V_n=-\dfrac{1}{2}M^n_{A} R_n^{A}+\dfrac{1}{4}L^n_{{A}{A'}}R_n^{A}R_n^{{A'}},
\label{VB}
\end{equation}
where $M^n_{A} $ is the $1 \times n(2n-1)$-dimensional mass matrix and
$L^n_{{A}{A'}}$ is a quartic coupling matrix with
$n(2n-1) \times n(2n-1)$ entries.  Evidently, for a U(1$)_Y$-invariant
$n$HDM potential, the first $n^2$ elements of $M^n_{A} $ and
$n^2 \times n^2$ elements of $L^n_{{A}{A'}}$ are only relevant, since
the other U(1$)_Y$-violating components vanish.  The general 2HDM and
3HDM potentials with their corresponding $M^n_{A} $ and
$L^n_{{A}{A^\prime}}$ are presented in Appendix~\ref{V2-3}.

The gauge-kinetic term of the $\bm{\Phi}_n$-multiplet is given by
\begin{equation}
T_n=\frac{1}{2}(D_{\mu}{\bm{\Phi}}_n)^{\dagger} (D^{\mu}{\bm{\Phi}}_n),
\end{equation}
where the covariant derivative in the ${\bm{\Phi}}_n$ space is
\begin{equation}
 D_{\mu}\ =\ \sigma^0 \otimes {\bf{1}}_n \otimes
          (\sigma^0 \partial^0_{\mu} +ig_w W_{\mu}^i \sigma^i /2)\:
+\: \sigma^3 \otimes {\bf{1}}_n \otimes i{g_Y\over 2} B_{\mu} \sigma^0.
\end{equation}
In the limit $g_Y \to 0$, the gauge-kinetic term $T_n$ is invariant
under $\text{Sp}(2n)/Z_2 \otimes \text{SU}(2)_L$ transformations of
the $\bm{\Phi}_n$-multiplet. In general, the maximal symmetry group
acting on the ${\bm{\Phi}}_n$-space in the $n$HDM potentials is
\begin{equation*}
\text{G}_{\mathsf{n\text{-HDM}}}^{{\bm{\Phi}}_n}=\text{Sp}(2n)/Z_2 \otimes \text{SU}(2)_L,
\end{equation*}
which leaves the local SU($2)_L$ gauge kinetic term of ${\bm{\Phi}}_n$
canonical.  The local SU(2$)_L$ group generators can be represented as
$\sigma^0 \otimes {\bf{1}}_n \otimes (\sigma^{1,2,3}/2)$, which commute
with all generators of the Sp($2n$) group.

Let us turn our attention to the Sp(2$n$) generators 
${K}^{{B}}_n$, with ${B =0,1,\dots,n(2n+1)-1}$. They satisfy the
important relation~\cite{Pilaftsis:2011ed},
\begin{equation}
C^{-1}{K}^{{B}}_n C=-K_n^{{{B}}*}=-({K}^{{B}}_n)^{\mathsf{T}},
\end{equation}
which implies that ${K}^{{B}}_n$ are C-odd.  The Sp($2n)$ generators
may conveniently be expressed in terms of double tensor products as~\cite{Georgi:1982jb}
\begin{gather}
{K}^{{B}}_n=\left( \sigma^0 \otimes t_A^b \otimes \sigma^0, \quad \sigma^i \otimes t_S^a 
\otimes \sigma^0 \right),
\end{gather}
where $t_S^a$ ($ t_A^b$) are the symmetric (anti-symmetric) generators
of the SU($n$) symmetry group. For instance, the 10 generators of
Sp(4) are~\cite{Pilaftsis:2011ed}
\begin{align}
&K^{0,1,3}_2=\dfrac{1}{2}\sigma^3 \otimes \sigma^{0,1,3} \otimes \sigma^0, \, \quad \quad \;\;\;
K^2_2=\dfrac{1}{2}\sigma^0 \otimes \sigma^2 \otimes \sigma^0,\nonumber \\
&K^{4,5,8}_2=\dfrac{1}{2}\sigma^1 \otimes \sigma^{0,3,1} \otimes \sigma^0, \quad \quad 
K^{6,7,9}_2=\dfrac{1}{2}\sigma^2 \otimes \sigma^{0,3,1} \otimes \sigma^0.
\end{align}
Likewise, the Sp(6) group has 21 generators, which are
\begin{align}
K^{0,1,2,3,4,5}_3&=\dfrac{1}{2}\sigma^3 \otimes G^{0,1,3,4,6,8}
                   \otimes \sigma^0, \qquad \qquad \quad \quad 
K^{6,7,8}_3=\dfrac{1}{2}\sigma^0 \otimes G^{2,5,7} \otimes \sigma^0,
 \nonumber \\
K^{9,10,11,12,13,14}_3&=\dfrac{1}{2}\sigma^1 \otimes G^{0,1,3,4,6,8} \otimes \sigma^0, \quad \quad K^{15,16,17,18,19,20}_3=\dfrac{1}{2}\sigma^2\otimes G^{0,1,3,4,6,8}\otimes \sigma^0.
\end{align}

It is interesting to state the Lie commutators between the
${\Sigma}^{A}_n$ and ${K}^{{B}}_n$ generators,
\begin{equation}
[{K}^{{B}}_n, {\Sigma}^{I}_n]=2if_n^{{B}IJ}{\Sigma}^{J}_n,
\label{commutation}
\end{equation}
where $I,J=1,\dots, n(2n-1)-1$ and $f_n^{{B}IJ}$ is a subset of the
structure constants of the SU(2$n$) group. Employing
(\ref{commutation}), we may define the Sp(2$n$) generators in the
bi-adjoint representation (i.e. the adjoint representation in the bilinear formalism) as
\begin{equation}
(T^{{B}}_{n})_{IJ}\ =\ -i f_n^{{B}IJ}\ =\ \text{Tr} \Big([ {\Sigma}^I_n, K^{{B}}_n]{\Sigma}^{J}_n\Big)\,. 
\label{TIJ}
\end{equation}
Note that the dimensionality of the bi-adjoint representation differs from the standard adjoint
representation. The former representation has
$(2n^2-n-1) \times (2n^2-n-1)$ dimensions, whereas the latter has $n(2n+1) \times n(2n+1)$ dimensions
~\cite{Hamermesh,Preskill:1980mz,Georgi:1999wka,Vergados}. 
The generators of $T^{{B}}_{n}$ for Sp(4) and Sp(6) corresponding to the 2HDM and
3HDM are presented in Appendix~\ref{sp-g}.

Knowing that Sp(2$n)$ is the maximal symmetry group allows us to
classify all SU(2)$_L$-preserving accidental symmetries of $n$HDM
potentials. These symmetries can be grouped into two categories: (i)
continuous symmetries and (ii) discrete symmetries (Abelian and
non-Abelian symmetry groups). In the next section, we demonstrate the
structure of continuous symmetries and prime invariants for building
$n$HDM potentials.

\section{Continuous Symmetries and Prime invariants}
\label{sec:cs}

The symplectic group Sp($2n)$ acts on the ${\bm{\Phi}}_n$-space, such
that the bilinear vector $R_{n}^{A}$ transforms in the bi-adjoint
representation of Sp(2$n$) defined in (\ref{commutation}) and
(\ref{TIJ}). It is therefore essential to consider the maximal
subgroups of Sp(2$n$).  Then, the accidental maximal subgroups would
be the combinations of smaller symplectic groups, such
as~\cite{Slansky:1981yr}
\begin{equation}
\text{Sp}(2n) \supset \text{Sp}(2p) \otimes \text{Sp}(2q),
\end{equation}
where $p+q=n$. The other maximal subgroup is
\begin{equation}
\text{Sp}(2n) \supset \text{SU}(n) \otimes \text{U}(1).
\end{equation}
The breaking pattern of SU($n$) in the maximal subgroups are
\begin{align}
\text{SU}(n) & \supset \text{SU}(p) \otimes \text{SU}(q) \otimes \text{U}(1), \\
& \supset \text{Sp}(2k), \\
& \supset \text{SO}(n),
\end{align}
with $p+q+1=n$ and $n\leq 2k$. The breaking pattern for SO($n$) is
\begin{align}
\text{SO}(n) & \supset \text{SO}(p) \otimes \text{SO}(q),
\end{align}
where $p+q=n$. 
Note that local isomorphisms should also be taken into account, such as
\begin{eqnarray}
&&\text{SO(3)} \cong \text{SU(2)} \cong \text{Sp(2)}, 
\nonumber \\
&&\text{SO(4)} \cong \text{SU(2)} \otimes \text{SU(2)}, 
\nonumber \\
&&\text{SO(5)} \cong \text{Sp(4)},
\nonumber \\
&&\text{SO(6)} \cong \text{SU(4)}~\text{etc}.
\nonumber
\end{eqnarray}

Following this procedure, it is straightforward to identify all
accidental continuous symmetries for $n$HDM potentials. For the
simplest case, i.e.~that of the 2HDM, the above breaking chain gives rise to the
following continuous symmetries \cite{Pilaftsis:2011ed}, 
\vspace{-0.2in}
\begin{align}
\begin{minipage}{0.5\textwidth}
\begin{eqnarray}
&&(a) \ \text{Sp(4)} \cong \text{SO(5)}, 
\nonumber \\[0.04in] 
&&(b) \ \text{Sp(2)} \otimes \text{Sp(2)} \ \cong \text{SO(4)},
\nonumber \\[0.04in] 
&&(c) \ \text{Sp(2})_{\text{}} \cong \text{SO(3)},
 \nonumber \\[0.04in] 
&&(d) \ \text{SU}(2)_{\text{HF}}\cong \text{O(3)}\otimes \text{O(2)}, 
 \nonumber
\end{eqnarray}
\end{minipage}
\begin{minipage}{0.5\textwidth}
\begin{eqnarray}
&&(e) \ \text{SO(2})_{\text{HF}}\ \cong Z_2\otimes [\text{O(2)}]^2, 
 \nonumber \\[0.04in] 
&&(f) \ \text{U(1})_{\text{PQ}}\otimes \text{Sp}(2)_{\text{}} \cong \text{O(2)}\otimes \text{O(3)}, 
 \nonumber \\[0.04in] 
&&(g) \  \text{U(1})_{\text{PQ}}\otimes \text{U(1})_Y
 \cong \text{O(2)}\otimes \text{O(2)}, 
  \nonumber \\[0.04in] 
  && 
 \nonumber
\end{eqnarray}
 \end{minipage}
 \label{s2hdm}
 \nonumber \\
\end{align}
where HF indicates Higgs Family symmetries that only involve the
elements of $\Phi=(\phi_1,\phi_2,\dots,\phi_n)^{\mathsf{T}}$ and not
their complex conjugates.  In the case of 3HDM, the maximal symmetry
is Sp(6), so in addition to all symmetries in (\ref{s2hdm}) we find
\vspace{-0.2in}
\begin{align}
\begin{minipage}{0.5\textwidth}
\begin{eqnarray}
&&(h) \ \text{Sp(6)},
\nonumber \\[0.04in] 
&&(i) \ \text{Sp(4)} \otimes \text{Sp(2)},
 \nonumber \\[0.04in] 
&& (j) \ \text{Sp(2)} \otimes \text{Sp(2)}\otimes \text{Sp(2)},
 \nonumber
 \end{eqnarray}
 \end{minipage}
\begin{minipage}{0.4\textwidth}
\begin{eqnarray}
&&(k) \ \text{SU(3)}\otimes \text{U(1)},
 \nonumber \\[0.04in] 
&&(l) \ \text{SO(3)}.
 \nonumber \\[0.04in] 
  && 
 \nonumber
\label{s3hdm}
\end{eqnarray}
 \end{minipage}
 \nonumber \\
\end{align}
We may now construct accidentally symmetric $n$HDM potentials in terms
of fundamental building blocks that respect the symmetries. To this end, we
introduce the invariants $S_n$, $D_n^2$ and~$T_n^2$. In detail, $S_n$
is defined as 
\begin{align}
   \label{eq:spn}
 S_n\ =\ {\bm{\Phi}}_n^\dagger{\bm{\Phi}}_n\;,
\end{align}
which is invariant under both the SU($n)_L\otimes$U(1)$_Y$ gauge group and Sp(2$n$).
Moreover, we define the SU(2)$_L$-covariant quantity $D^a_n$ in the HF
space as
\begin{equation}
    \label{su}
D_n^a\ =\ {{\Phi}}^\dagger \sigma^a {{\Phi}}\; . 
\end{equation}
Under an SU(2)$_L$ gauge transformation,
$D^a_n \to D_n^{\prime a}= O^{ab}D_n^b$, where $O\in$~SO(3).  Hence,
the quadratic quantity $D_n^2\equiv D^a_n D^a_n$ is both gauge and SU($n$)
invariant. Finally, we define the auxiliary quantity $T_n$ in the HF
space as
\begin{equation}
T_n\ =\ \Phi \Phi^{\mathsf{T}}\;,
\end{equation}
which transforms as a triplet under SU(2)$_L$,
i.e.~$T_n \to T_n'=U_L T_n U_L^{\mathsf{T}}$.  As a consequence, a
proper prime invariant may be defined as
$T_n^2\equiv \text{Tr}(TT^*)$, which is also both gauge and SO($n$)
invariant.

In addition to the above maximal prime invariants, it is useful to
define minimal invariants. For instance, prime invariants that respect Sp(2) can be derived from the doublets $\begin{pmatrix}
  \phi_i \\ i \sigma^2 \phi_i^*\end{pmatrix}$ and $\begin{pmatrix}
  \phi_i \\ 
  i \sigma^2 \phi_j^* \end{pmatrix}$,
\begin{align}
&S_{ij}=\begin{pmatrix} \phi_i^{\dagger}, -i
  \phi_i^{\mathsf{T}}\sigma^2 \end{pmatrix}\begin{pmatrix} \phi_j \\ 
i \sigma^2 \phi_j^* \end{pmatrix}=\phi_i^\dagger \phi_j+\phi_j^\dagger \phi_i, \nonumber \\
&S^{\prime}_{ij}=\begin{pmatrix} \phi_i^{\dagger}, -i
  \phi_j^{\mathsf{T}}\sigma^2 \end{pmatrix}\begin{pmatrix} \phi_i \\ 
i \sigma^2 \phi_j^* \end{pmatrix}=\phi_i^\dagger \phi_i+\phi_j^\dagger \phi_j.
\end{align}

Likewise, to obtain minimal SU(2)$\otimes$U(1) invariants from
$\begin{pmatrix} \phi_i \\ \phi_j \end{pmatrix}$ and $\begin{pmatrix} \phi_i \\
  i \sigma^2 \phi_j^* \end{pmatrix}$, we define
\begin{align}
& D_{ij}^a =\begin{pmatrix} \phi_i^{\dagger},
  \phi_j^{\dagger} \end{pmatrix} \sigma^a \begin{pmatrix} \phi_i \\ 
 \phi_j \end{pmatrix} =\phi_i^{\dagger} \sigma^a \phi_i
  +\phi_j^{\dagger} \sigma^a \phi_j =D_{ji}^a, \nonumber \\ 
& D^{\prime a}_{ij}=\begin{pmatrix} \phi_i^{\dagger}, -i
  \phi_j^{\mathsf{T}}\sigma^2 \end{pmatrix} \sigma^a \begin{pmatrix}
  \phi_i \\ 
i \sigma^2 \phi_j^* \end{pmatrix}=\phi_i^{\dagger} \sigma^a \phi_i -
  \phi_j^{\dagger} \sigma^a \phi_j=-D_{ji}^{\prime a}, 
\end{align}
where the identity $\sigma^2 \sigma^a \sigma^2=-(\sigma^a){}^{\mathsf{T}}$ has been used.

By analogy, to construct an SO(2)-invariant expression from the doublet
$\begin{pmatrix} \phi_i \\ \phi_j \end{pmatrix}$, we may use quantities such as
\begin{align}
T_{ij}=\phi_i \phi_i^{\mathsf{T}} +\phi_j \phi_j^{\mathsf{T}} =T_{ji}.
\end{align}
Moreover, extra prime invariants can be constructed from
$\begin{pmatrix}\phi_i \\ i \sigma^2 \phi_i^* \end{pmatrix}$ and
$\begin{pmatrix} \phi_j \\ i \sigma^2 \phi_j^* \end{pmatrix}$, e.g. 
\begin{align}
S_{ij} &=\phi_i^{\dagger} \phi_j +\underbrace{{ (i \sigma^2 \phi_i^*)}^{\dagger}{ (i \sigma^2 \phi_j^*)}}_{\phi_i^{\mathsf{T}} \phi_j^*}=\phi_i^{\dagger} \phi_j +\phi_j^{\dagger} \phi_i,
\\
iD_{ij}^{\prime\prime a} &=\phi_i^{\dagger}\sigma^a \phi_j + {(i \sigma^2 \phi_i^*)}^{\dagger} \sigma^a (i \sigma^2 \phi_j^*) =\phi_i^{\dagger} \sigma^a \phi_j -\phi_j^{\dagger} \sigma^a \phi_i.
\end{align}
Note that $D_{ij}^{\prime\prime a} D_{ij}^{\prime\prime a}$ depends on $S_{ij}$ and $ S_{ii,jj}$, since
\begin{align}
-{D_{ij}^{\prime\prime 2}}=S_{ij}^2 - (\phi_i^{\dagger} \phi_j)(\phi_j^{\dagger} \phi_j)=S_{ij}^2 -S_{ii} S_{jj}.
\end{align}
Observe that $T_{ij}$ and $S_{ij}$ are not invariant under phase
re-parametrizations, $\phi_i \to e^{i \alpha_i} \phi_i$, and so they
need to be appropriately combined with their complex conjugates.

We are now able to build a symmetric scalar potential
$V_{\rm sym}$ in terms of prime invariants as follows:
\begin{align}
V_{\rm sym}\ =\ -\mu ^2 S_n\: +\:\lambda_S S_n^2\: +\: \lambda_D
  D_n^2\: +\:
  \lambda_T T_n^2\; .
\end{align}
Obviously, the simplest form of the $n$HDM potentials obeys the
maximal symmetry~Sp(2$n$), which has the same form as the SM
potential,
$$
V_{\text{SM}}\ =\ -\mu^2 \left( \phi^{\dagger}\phi \right)\:  +
 \:  \lambda \left( \phi^{\dagger}\phi\right)^2,
$$
with a single mass term and a single quartic coupling. For example,
the 2HDM Sp(4)-invariant potential, the so-called MS-2HDM is
given by~\cite{Dev:2014yca}
\begin{equation}
    \label{eq:MS2HDM}
V_{\text{MS-2HDM}}\ =\ - \mu_1^2 \left( \left| \phi_1 \right|^2\: +\: \left| \phi_2 \right|^2 \right) 
 + \lambda_1 \left( \left| \phi_1 \right|^2 + \left| \phi_2 \right|^2 
   \right)^2\; ,
\end{equation}
with the obvious relations among the parameters,
\begin{align}
\mu_1^2 = \mu_2^2, \quad m_{12}^2 = 0,\quad
2\lambda_2 = 2\lambda_1= \lambda_3, \quad
\lambda_4 = \text{Re}(\lambda_5) = \lambda_6 = \lambda_7 = 0.
\end{align}
Note that the functional form of the potential in~\eqref{eq:MS2HDM} 
is $V= V[S_2] = V[S_{11}+S_{22}]$.

Similarly, the 3HDM potential invariant under Sp(6)${ /Z_2}$ will be a
function of the symmetric block $S_3 = S_{11}+S_{22}+S_{33}$, i.e.
\begin{equation}
V_{\text{MS-3HDM}}\ =\ - \mu_1^2 \left( \left| \phi_1 \right|^2\: +\:
  \left| \phi_2 \right|^2\: +\: \left| \phi_3 \right|^2 \right)  
 + \lambda_{11} \left( \left| \phi_1 \right|^2 + \left| \phi_2 \right|^2\: +\: \left| \phi_3 \right|^2
   \right)^2, 
\end{equation}
where the non-zero parameters are
\begin{align}
\mu_1=\mu_2=\mu_3,\quad \lambda_{11}=\lambda_{22}=\lambda_{33}=2\lambda_{1122}=2\lambda_{1133}=2\lambda_{2233}.
\nonumber
\end{align}
Remarkably, the MS-$n$HDM potentials
obey naturally the SM-alignment constraints and all quartic couplings
of the MS-$n$HDM potential can vanish
simultaneously~\cite{Dev:2014yca,Darvishi:2019ltl}.

Another example is the SU(3)$\otimes$U(1)-invariant 3HDM
potential. The corresponding symmetric blocks are $S_3 =
S_{11}+S_{22}+S_{33}$ and $D^2_3 = D_{12}^2+D_{13}^2+D_{23}^2$, given
in~\eqref{eq:spn} and~\eqref{su}, respectively.
Therefore, the
SU(3)$\otimes$U(1)-invariant 3HDM potential takes on the form
\begin{align}
V_{\text{SU(3)}\otimes \text{U(1)}}&= - \mu_1^2 \left( |\phi_1|^2\:
                                     +\: |\phi_2|^2\: +\: | \phi_3 |^2
                                     \right)  
 + \lambda_{11} \left( |\phi_1|^4 + |\phi_2|^4\: +\: |\phi_3|^4
   \right)\
\nonumber \\ &
+ \lambda_{1122} \left( |\phi_1|^2 |\phi_2|^2 
+ |\phi_1|^2  |\phi_3 |^2 
 + |\phi_2|^2 \,|\phi_3|^2 \right)
\nonumber \\ &
+ (2 \lambda_{11}-\lambda_{1122}) \left( |\phi_1^{\dagger} \phi_2 |^2
+ |\phi_1^{\dagger} \phi_3|^2
+ |\phi_2^{\dagger} \phi_3|^2\right),
\end{align}
with the following relations between the parameters:
\begin{align}
&\mu_1=\mu_2=\mu_3,\quad \lambda_{11}=\lambda_{22}=\lambda_{33},\quad \lambda_{1122}=\lambda_{1133}=\lambda_{2233},\nonumber \\
&\lambda_{1221}=\lambda_{1331}=\lambda_{2332}=2 \lambda_{11}-\lambda_{1122}.\nonumber 
 \end{align}
 
 This method can be applied to all continuous symmetries of 
 $n$HDM potentials. We present all explicit symmetric blocks under all
 continuous symmetries for the 2HDM and the 3HDM potentials in Appendix
 \ref{C}. The complete list of  accidental symmetries for the 2HDM
 and 3HDM potentials, along with the relations among the non-zero
 parameters, are exhibited in Tables~\ref{tab1} and~\ref{tab2}.

\section{Discrete symmetries and Irreducible representations}
\label{sec:ds}

As discussed in Section \ref{sec:s1}, Sp(2$n)$ is the maximal symmetry
group for $n$HDM potentials. This will help us to classify all
SU(2)$_L$-preserving accidental symmetries of such potentials. In
addition to continuous symmetries shown in Section \ref{sec:cs}, there
are also discrete symmetries as subgroups of continuous symmetries.
Known examples of this type are the Standard CP symmetry, the Cyclic
discrete group $Z_n$, the Permutation group $S_n$, or a product of
them possibly combining with continuous symmetries.  In general, these
discrete symmetries can be grouped into Abelian and non-Abelian
symmetry groups.  In this section, we discuss all possible and known
discrete symmetries for $n$HDM potentials, including our approach to
build $n$HDM potentials by employing irreducible representations of
discrete symmetry groups.

\subsection{Abelian Discrete Symmetries}

To start with, let us first consider the Abelian discrete symmetry
groups~\cite{Humphreys,Beardon} 
\begin{equation}
\quad Z_2, \quad Z_3, \quad Z_4, \quad Z_2 \times Z_2, \quad \dots, \quad Z_n, \quad \dots,
\end{equation}
where $Z_n=\{1, \omega, \dots, \omega^{(n-1)}\}$ with $\omega^n=1$.
Note that iff $n$ and $m$ have no common prime factor, the product
$Z_n \times Z_m$ is identical to $Z_{n\times m}$. These discrete
symmetries can be imposed to restrict the independent theoretical
parameters of the model. For example, in the 2HDM, the $Z_2$ symmetry is
invoked to avoid flavour changing neutral currents~\cite{Glashow:1976nt} or to ensure the
stability of dark matter~\cite{Silveira:1985rk}.

In addition to these discrete symmetries, there are Generalized CP
(GCP) transformations defined as,
\begin{equation}
\text{GCP}[{{\phi}}_i]=G_{ij} {{\phi}}_{j}^*,
\end{equation}
with $G_{ij} \in \text{SU(}n)\times$U(1), where U(1) can always be
eliminated by U(1)$_Y$ transformation. The GCP transformations realize
different types of CP symmetry. For example, in the case of the 2HDM
(3HDM), there are two types of CP symmetries: (i)~standard CP or CP1:
${{{\phi}}_{1,2}\, (\phi_3)\to {{\phi}}_{1,2} ^*\,(\phi^*_3)}$ and (ii)~non-standard CP or CP2:
${{\phi}}_1 \to {{\phi}}_2^*$, ${{\phi}}_2 \to -{{\phi}}_1^*$
(${{{\phi}}_3 \to
  {{\phi}}_3^*}$)~\cite{Ferreira:2009wh,Ferreira:2010yh}\footnote{In
  some of the 
  earlier literature, there has been a third class of CP symmetries
  called CP3 which relies on SO(2)$_{\text{HF}}$ (or
  SO(3)$_{\text{HF}}$ for a 3HDM potential) and the standard CP
  symmetry. However, continuous SO$(n)_{\text{HF}}$ symmetries lead
  automatically to CP-invariant $n$HDM potentials, without introducing
  further restrictions on the theoretical parameters. In other words,
  CP1 is an emergent symmetry that results from SO$(n)_{\text{HF}}$,
  so CP3 should not be regarded as a new CP symmetry.}.  In general,
without continuous group factors, the discrete symmetries for the 2HDM
are CP1, CP2 and $Z_2$. The generators of these discrete symmetries
can be expressed in terms of double tensor products as
\begin{eqnarray}
\Delta_{\text{CP1}}&=&\sigma^2 \otimes \sigma^0 \otimes \sigma^2,
\\
\Delta_{\text{CP2}}&=&i \sigma^2 \otimes \sigma^2 \otimes \sigma^2,
\\
\Delta_{Z_2}&=&\sigma^0 \otimes \sigma^3 \otimes \sigma^0.
\end{eqnarray}
In the bilinear $R_2^{A}$-space, the transformation matrices (or the generating group elements) associated with CP1, CP2 and $Z_2$ discrete symmetries are given by
\begin{align}
D_{\text{CP1}}&=\text{diag} ({\bf{1,-1,1}},1,-1), \\
D_{\text{CP2}}&=\text{diag} ({\bf{-1,-1,-1}},1,-1),\\
D_{Z_2}&=\text{diag} ({\bf{-1,-1,1}},-1,-1).
\end{align}
where the U(1)$_Y$-conserving elements are denoted in boldface.

Let us turn our attention to the 3HDM potential. In this case, the corresponding CP1 discrete symmetry can be represented as 
\begin{equation}
\Delta_{\text{CP1}}=\sigma^2 \otimes G^0 \otimes \sigma^2,
\end{equation}
resulting in the following transformation matrix in the $R_3^{A}$-space,
\begin{align}
\small
D_{\text{CP1}} &=\text{diag} ({\bf{1,1,1,-1,-1,-1,1,1}},1,1,1,-1,-1,-1).
\end{align}
On the other hand, CP2 discrete symmetries may be given by
\begin{align}
 \Delta_{\text{CP2}}&=\begin{pmatrix}0&-i \bar{G}^2 \otimes \sigma^2 \\ i\bar{G}^{2^*} \otimes \sigma^2&0 \end{pmatrix}, 
\end{align}
with~\cite{Ivanov:2015mwl}
\begin{align}
\bar{G}^{2} ={\begin{pmatrix}0&-1&0\\ 1&0&0\\0&0& e^{i \alpha} \end{pmatrix}},
\end{align}
where the phase $e^{i \alpha}$ is an arbitrary phase factor. This
results in the following transformation matrix in the bilinear
$R_3^{A}$-space (dots stand for zero elements)
\begin{align}
&\setlength{\arraycolsep}{0.3pt}
\renewcommand{\arraystretch}{0.01}
D_{\text{CP2}}=\left(\scriptsize
\begin{array}{cccccccccccccc}
\bf{-1 }&. & . & . & . & . & . &. & . & . & . & . & . & . \\
. &. &\bm{-\cos{\bm\alpha}}  & . & . &\bm{\sin{\bm\alpha}} & . & . & . & . & . & . & . & .\\
. &\bm{\cos{\bm\alpha}}  & . & .  &\bm{-\sin{\bm\alpha}} & . & . & . & . & . & . & . & . & . \\
. &. & . & \bf{-1} & . & . & . & . & . & . & . & . & . & . \\
. &. &\bm{\sin{\bm\alpha}} & . & . &\bm{\cos{\bm\alpha}} & . & . & . & . & . & . & . & . \\
. &\bm{-\sin{\bm\alpha}}& . & . & \bm{-\cos{\bm\alpha}}  & . & . & . & . & . & . & . & . & . \\
. &. & . & . & . & . &\bf{-1 }& . & . & . & . & . & . & . \\
. &. & . & . & . & . & . &\bf{1} & . & . & . & . & . & . \\
. & . &. & . &. & . & . & . & 1 & . & . &. & . & . \\
 . & . & . & . & . & . & . & . & . & . &-\cos\alpha & . &. &\sin\alpha \\
 . & . & . & . & . & . & . & . & . &\cos\alpha & . & . &-\sin\alpha&.\\
 . & . & . & . & . & . & . & . & .& . & . &-1 & . & . \\
 . & . & . & . & . & . & . & . & . & .&-\sin\alpha & . & . & \cos\alpha \\
 . & . & . & . & . & . & . & . & . & \sin\alpha& .& . & -\cos\alpha & . \\
\end{array}
\right).
\end{align}
Note that the CP2 transformation matrix $D_{\text{CP2}}$ in  the bilinear
$R_3^{A}$-space is non-diagonal, contrary to the 2HDM case. We must
remark here that in the case of 2HDM,
$\Delta_{\text{CP2}}^2 =-\bf{1}_8\neq \bf{1}_8$ and
$\Delta_{\text{CP2}}^4 =\bf{1}_8$ and in the bilinear space
$D_{\text{CP2}}^2 =\bf{1}_6$. However, in the case of the 3HDM, we have
$\Delta_{\text{CP2}}^2 \neq -\bf{1}_{12}$ and
$\Delta_{\text{CP2}}^4= \bf{1}_{12}$ and
$D_{\text{CP2}}^4 =\bf{1}_{14}$, in agreement with a property termed
CP4 in~\cite{Ivanov:2015mwl}. Without loss of generality, we
set $\alpha=0$.

In addition, there are several Abelian discrete symmetries for the 3HDM potential~\cite{Ivanov:2011ae},~i.e.
\begin{equation}
\quad Z_2, \quad Z^{\prime}_2, \quad Z_3, \quad Z_4, \quad Z_2 \times Z_2.
\end{equation}
The generators of these Abelian discrete symmetries are given by
\begin{align}
\Delta_{{Z_2}}&=\sigma^0 \otimes {z_2} \otimes \sigma^0,
\qquad \qquad \quad \,\,
\Delta_{{Z^{\prime}_2}}=\sigma^0 \otimes {z^{\prime}_2} \otimes \sigma^0,
\nonumber \\
\Delta_{{Z_3}}&=\text{diag}[{z_3} \otimes \sigma^0,{z_3}^* \otimes \sigma^0],
\quad
\Delta_{{Z_4}}=\text{diag}[{z_4} \otimes \sigma^0,{z_4}^* \otimes \sigma^0],
\label{22p34}
\end{align}
where
\begin{align}
{z_2}&=\text{diag}[1,-1,1],
\quad
{z^{\prime}_2}=\text{diag}[1,-1,-1],
\nonumber \\
{z_3}&=\text{diag}[\omega^2,\omega,1],
\quad
{z_4}=\text{diag}[i,-i,1],
\end{align}
with $\omega=e^{i2\pi/3}$.  In the bilinear $R_3^{A}$-space, as a
result of flipping the sign of the one or two doublets, the following
diagonal transformation matrices for the discrete symmetries $Z_2$ and
$Z_2^{\prime}$ may be derived:
\begin{align}
\small
D_{{Z_2}}&=\text{diag} ({ {\bm{-1}}, {\bm{1}}, {\bm{-1}}, {\bm{-1}}, {\bm{1}}, {\bm{-1}}, {\bm{1}}, {\bm{1}}},-1,1,-1,-1,1,-1), \\
D_{{Z^{\prime}_2}}&=\text{diag} ({{\bm{-1}}, {\bm{-1}}, {\bm{1}}, {\bm{-1}}, {\bm{-1}}, {\bm{1}}, {\bm{1}}, {\bm{1}}},-1,-1,1,-1,-1,1).
\label{z22}
\end{align}
In the same way, the transformation matrices for $Z_3$ and $Z_4$ may
be represented by the non-diagonal matrices
\begin{align}
&\setlength{\arraycolsep}{1pt}
\renewcommand{\arraystretch}{0.9}
D_{{Z_3}}={1\over 2}\left(\scriptsize
\begin{array}{cccccccccccccc}
 {\bm{-1}} &. & . &{\bm{\sqrt{3}}} & . & . & . & . & . & . & . & . & . & . \\
. & {\bm{-1}}& . & . & {-\bm{\sqrt{3}}}& . & . & . & . & . & . & . & . & .\\
. & . & {\bm{-1}} & .  & . &{\bm{\sqrt{3}}} & . & . & . & . & . & . & . & . \\
{-\bm{\sqrt{3}}} &. & . & {\bm{-1}} & . & . & . & . & . & . & . & . & . & . \\
. &{\bm{\sqrt{3}}}& . & . & {\bm{-1}} & . & . & . & . & . & . & . & . & . \\
. &. &{-\bm{\sqrt{3}}}& . &. & {\bm{-1}}& . & . & . & . & . & . & . & . \\
. &. & . & . & . & . & {\bm{2}} & . & . & . & . & . & . & . \\
. &. & . & . & . & . & . & {\bm{2}}& . & . & . & . & . & . \\
. & . &. & . &. & . & . & . & 2 & . & . &. & . & . \\
 . & . & . & . & . & . & . & . & . & -1 &. & . &\sqrt{3} &. \\
 . & . & . & . & . & . & . & . & . &. & -1 & . & . &-\sqrt{3}\\
 . & . & . & . & . & . & . & . & .& . & . &2 & . & . \\
 . & . & . & . & . & . & . & . & . &-\sqrt{3}& . & . & -1 & . \\
 . & . & . & . & . & . & . & . & . & . &\sqrt{3}& . & . & -1\\
\end{array}
\right),
\end{align}
\begin{align}
&\setlength{\arraycolsep}{3pt}
\renewcommand{\arraystretch}{0.6}
D_{{Z_4}}={1\over 2}\left(\scriptsize
\begin{array}{cccccccccccccc}
 {\bm{-1}} &. & . &. & . & . & . & . & . & . & . & . & . & . \\
. &. & . & . & {\bm{1}}& . & . & . & . & . & . & . & . & .\\
. & . &. & .  & . & {\bm{-1}} & . & . & . & . & . & . & . & . \\
 &. & . & {\bm{-1}}& . & . & . & . & . & . & . & . & . & . \\
. & {\bm{-1}}& . & . &. & . & . & . & . & . & . & . & . & . \\
. &. & {\bm{1}} & . &. & .& . & . & . & . & . & . & . & . \\
. &. & . & . & . & . & {\bm{1}} & . & . & . & . & . & . & . \\
. &. & . & . & . & . & . & {\bm{1}}& . & . & . & . & . & . \\
. & . &. & . &. & . & . & . & 1 & . & . &. & . & . \\
 . & . & . & . & . & . & . & . & . & . &. & . &-1 &. \\
 . & . & . & . & . & . & . & . & . &. & . & . & . &1\\
 . & . & . & . & . & . & . & . & .& . & . &1 & . & . \\
 . & . & . & . & . & . & . & . & . &1& . & . & . & . \\
 . & . & . & . & . & . & . & . & . & . &-1& . & . &.\\
\end{array}
\right).
\end{align}

Now, with the help of the $D$-transformation matrices, we can
construct accidentally symmetric $n$HDM potentials. For example, in
the case of $Z_4$ symmetry, applying the transformation matrix
$D_{{Z_4}}$ on the U(1)$_Y$-conserving elements of $R_3^{A}$ yields
\begin{align}
&R_3^1\to -R_3^1, &R_3^5\to -R_3^2, 
\nonumber \\
&R_3^2\to +R_3^5,  &R_3^6\to +R_3^3,
\nonumber \\
&R_3^3\to -R_3^6, &R_3^7\to +R_3^7,
\nonumber \\
&R_3^4\to -R_3^4, &R_3^8\to +R_3^8.
\end{align}
Therefore, all possible combinations $R_3^i R_3^j$ ($i,j=1,\dots, 8$) that
respect the $Z_4$ symmetry may be obtained as
\begin{align}
&(R_3^1)^2, \quad (R_3^4)^2, \quad R_3^1 R_3^4, \quad
(R_3^7)^2, \quad (R_3^8)^2, \quad R_3^7 R_3^8,
\nonumber \\
&R^{1,4}\big[(R^2_3 \pm R_3^3) - (R_3^5 \mp R_3^6)\big], \quad
(R_3^5 \pm R_3^6)^2+(R_3^2 \mp R_3^3)^2\;.
\end{align}
These combinations lead to the following $Z_4$-invariant 3HDM potential:
\begin{eqnarray}
V_{Z_4} =&-& \mu_1^2 \left| \phi_1 \right|^2
- \mu_2^2 \left| \phi_2\right|^2
- \mu_3^2 \left| \phi_3\right|^2
\nonumber \\
&+& \lambda_{11} \left| \phi_1 \right|^4
+ \lambda_{22} \left| \phi_2\right|^4
+ \lambda_{33} \left| \phi_3\right|^4
\nonumber \\
&+& \lambda_{1122} \left| \phi_1\right|^2 \left|\phi_2\right|^2
+ \lambda_{1133}  \left| \phi_1\right|^2 \left|\phi_3\right|^2
+ \lambda_{2233}  \left| \phi_2\right|^2 \left|\phi_3\right|^2
\nonumber \\
&+& \lambda_{1221} |\phi_1^{\dagger} \phi_2 |^2
+ \lambda_{1331} |\phi_1^{\dagger} \phi_3 |^2
+ \lambda_{2332} |\phi_2^{\dagger} \phi_3 |^2
\nonumber \\ &+& {\lambda_{1323} } (\phi_1^{\dagger} \phi_3) (\phi_2^{\dagger} \phi_3) 
+ {\lambda_{1323}^* (\phi_3^{\dagger} \phi_1) (\phi_3^{\dagger} \phi_2)} \nonumber \\
&+& 
{\lambda_{1212} \over 2} (\phi_1^{\dagger} \phi_2)^2 +
    {\lambda_{1212}^* \over 2} (\phi_2^{\dagger} \phi_1)^2, 
\label{z4}
\end{eqnarray}
where the complex phase of $\lambda_{1212}$ can be rotated away by a field redefinition.

In the same fashion, we can use this procedure to construct $n$HDM
potentials invariant under all Abelian discrete
symmetries. Tables~\ref{tab1} and~\ref{tab2} give the parameter
relations of the 2HDM and 3HDM potentials constrained by these
symmetries.

\subsection{Non-Abelian Discrete Symmetries}

Non-Abelian discrete symmetries constitute another class of discrete
symmetries, which may be thought of as combinations of Abelian
discrete symmetries. The most familiar non-Abelian groups can be
summarized as follows~\cite{Frampton:1994rk,Ishimori:2012zz}:

(i) \textit{Permutation group} $S_N$. The best known non-Abelian
discrete groups are the permutation groups. The order of this group is
$N!$. An $S_2$ group is an Abelian symmetry group and consists of a
permutation in the form $(x_1, x_2)\to (x_2, x_1)$. Thus, the lowest
order non-Abelian group is $S_3$.

(ii) \textit{Alternating group} $A_N$. This group consists of only
even permutations of $S_N$ and thus, its order is $N!/2$. The smallest
non-Abelian group of this class is $A_4$ since $A_3 \cong Z_3$.

(iii) \textit{Dihedral group} $D_N$. This group is also denoted as
$\Delta(2N)$ and its order is $2N$. The $D_N$ group is isomorphic to
$Z_N\rtimes Z_2$ that consists of the cyclic rotation $Z_N$ and its
reflections. Note that $D_3 \cong S_3$.

(iv) \textit{Binary Dihedral group} $Q_{2N}$. This group, which is
also called \textit{Quaternion group}, is a double cover of $D_N$
symmetry group and its order is 4$N$.

(v) \textit{Tetrahedral group} $T_N$. This group is of order $3N$ and
isomorphic to $Z_N\rtimes Z_3$, where $N$ is any prime number. The
smallest non-Abelian discrete symmetry of this type is $T_7$. This
would imply that a $T_N$-symmetric $n$HDM potential will also be
symmetric under $Z_7$.
 
(vi) \textit{Dihedral-like groups}. These generic groups obey the
following isomorphisms:
\begin{align}
&\Sigma(2 N^2)\cong (Z_N \times {Z'}_N) \rtimes Z_2, &\quad
                                                       \Delta(3N^2)
                                                       \cong (Z_N
                                                       \times {Z'}_N)
                                                       \rtimes Z_3,
                                                       \nonumber \\ 
& \Sigma(3 N^3)\cong (Z_N \times {Z'}_N \times {Z''}_N) \rtimes Z_3,
                                                     &\quad
                                                       \Delta(6N^2)
                                                       \cong
                                                       (Z_N\times
                                                       {Z'}_N) \rtimes
                                                       S_3. 
\end{align}
The simplest groups of this type are $\Sigma(2) \cong Z_2$, $\Delta(6)
\cong S_3$, $\Delta(24) \cong S_4$, $\Delta(12) \cong A_4$, and
$\Sigma(24) \cong Z_2 \times \Delta(12)$. 

({vii}) \textit{Crystal-like groups} $\Sigma (M \phi)$, with
$\phi=1,2,3$. These groups are of order $M$ and are given by
\begin{align}
&\Sigma (60\phi),\quad \Sigma (168 \phi ),\quad \Sigma (36\phi), \quad \Sigma (72\phi), \quad \Sigma (216\phi), \quad \Sigma (360\phi),
\end{align}
where $\Sigma (60)\cong A_5$ and $\Sigma (216)\cong A_4$.

The decomposition of tensor products of 3-dimensional irreducible representations of these groups are given in Appendix \ref{D}. Note that imposing many of these symmetry groups lead to identical potentials or to potentials that are invariant under continuous symmetries. These non-Abelian discrete symmetries can be the symmetry of $n$HDM potentials for sufficiently large $n$, as discussed in Section~\ref{sec:cs}.

In the case of the 3HDM, the complete list of non-Abelian discrete
symmetries has been reported
in~\cite{Ishimori:2012zz,Ivanov:2012fp,Keus:2013hya,Ivanov:2014doa,deMedeirosVarzielas:2019rrp}. There
are two non-Abelian discrete symmetries as subgroups of SO(3), namely
$D_3$ and $D_4$.  In addition, there are non-Abelian discrete
symmetries as subgroups of SU(3),
\begin{equation}
A_4, \, S_4, \, \left\{ \Sigma(18), \, \Delta(27), \, \Delta(54) \right\}, \, \Sigma(36), 
\end{equation}
where the symmetry groups stacked in curly brackets produce identical potentials. 
Here, we discuss the cases $D_3$ and $A_4$, while the description of the rest of these types of symmetries for the 3HDM potential may be found in Appendix \ref{D}.

Let us start with the smallest non-Abelian discrete group $D_3 \cong S_3$. The irreducible representations of the $D_3$ symmetry group can be expressed by two singlets, \textbf{1} and \textbf{1$^{\prime}$}, and one doublet \textbf{2}. The $2\otimes 2$ tensor product of this group decomposes as
\begin{eqnarray}
D_3: \quad \textbf{2} \otimes \textbf{2}=\textbf{1} \oplus \textbf{1$^{\prime}$} \oplus \textbf{2}.
\end{eqnarray}
Moreover, the generators of $D_3$ discrete symmetry group should
satisfy the conditions: ${g_1^3=1}$, ${g_2^2=1}$ and
$g_1 \cdot g_2 = g_2 \cdot (g_1 \cdot g_1)$.  Thus, two generators of
$D_3$ in terms of double tensor products are given by
\begin{equation}
\Delta_{D_3}^1=\text{diag}[{g_1} \otimes \sigma^0,{g_1}^* \otimes \sigma^0], \quad
\Delta_{D_3}^2=\sigma^0 \otimes {g_2} \otimes \sigma^0,
\end{equation}
where
\begin{eqnarray} \small
g_1 = 
	\begin{pmatrix}
	\omega^2 & 0 & 0 \\
	0 & \omega & 0 \\
	0 & 0 & 1
	\end{pmatrix},
\quad
g_2 = 
	\begin{pmatrix}
	0 & -1 & 0 \\
	-1 & 0 & 0 \\
	0 & 0 & 1
	\end{pmatrix}.
\end{eqnarray}
Imposing $D_3$ on the U(1)-conserving part of the $R^{A}_3$ vector will lead to the following linear decomposition: \vspace{-0.2in}
\begin{flalign}
\begin{minipage}{0.3\textwidth}
\begin{eqnarray} 
\textbf{1} &:& \;
	\begin{pmatrix}
	R_3^0
	\end{pmatrix},
\nonumber \\
\textbf{1$^{\prime}$} &:& \;
	\begin{pmatrix}
	R_3^7
	\end{pmatrix},
\nonumber \\
	\textbf{1$^{\prime\prime}$} &:& \;
	\begin{pmatrix}
	R_3^8
	\end{pmatrix},
	\nonumber
	\end{eqnarray}
\end{minipage}
\begin{minipage}{0.3\textwidth}
\begin{eqnarray}
\textbf{2} &:& \;
	\begin{pmatrix}
	R_3^1 \\ 
	R_3^4
	\end{pmatrix}.
\nonumber \\
	\textbf{2$^{\prime}$} &:& \;
	\begin{pmatrix} 
	-R_3^2-R_3^3 \\
	R_3^5-R_3^6
	\end{pmatrix},
\nonumber
\end{eqnarray}
\end{minipage} 
\begin{minipage}{0.3\textwidth}
\begin{eqnarray}
\textbf{2$^{\prime\prime}$ } &:& \;
	\begin{pmatrix} 
	R_3^2+ R_3^6 \\
	R_3^3+ R_3^5
	\end{pmatrix},
\nonumber \\
\textbf{2$^{\prime\prime\prime}$} &:& \;
	\begin{pmatrix} 
	R_3^2-R_3^6 \\
	R_3^3-R_3^5
	\end{pmatrix}.
\nonumber
\end{eqnarray}
\end{minipage} 
\label{DS3}
\end{flalign}
Note that in the bilinear space, there are three singlets \textbf{1},
\textbf{1$^{\prime}$} and \textbf{1$^{\prime \prime}$}, and two
doublets, \textbf{2} and \textbf{2$^{\prime}$}.  Thus, given the
irreducible representations in~\eqref{DS3}, we may parametrize the
$D_3$-invariant 3HDM potential as follows:
\begin{eqnarray} \small
V_{D_3} = &&
-M_1\,{\bf{ 1}} -M_2\,{\bf {1^{\prime\prime}}} 
	 + \Lambda_{0}\,{\bf{1^{\mathsf{2}}}}
    +\Lambda_{1}\,{\bf{{1^{\prime}}^ \mathsf{2}}}
    +\Lambda_{2}\,{\bf{{1^{\prime \prime}}^ \mathsf{2}}}
  \nonumber   \\
   &&
   +\Lambda_{3}\,{\bf{1}}\cdot{ \bf{1^{\prime\prime}}}
    +\Lambda_{4}\, {\bf{{{2^{\mathsf{T}}}}\cdot \bf{2}}}
   +\Lambda_{5}\, {\bf{{{2^{\prime}}^\mathsf{T}}\cdot \bf{2^{\prime}}}}
    \nonumber   \\
   &&
    +\Lambda_{6}\, {\bf{{{2}^\mathsf{T}}\cdot \bf{2^{\prime}}}}
   +\Lambda_{7}\, {\bf{{{2^{\prime \prime}}^\mathsf{T}}\cdot \bf{2^{\prime \prime}}}}
    +\Lambda_{8}\, {\bf{{{2^{\prime \prime\prime}}^\mathsf{T}}\cdot \bf{2^{\prime\prime \prime}}}}.
\end{eqnarray}
This can be rewritten as
\begin{eqnarray} \small
V_{D_3} =
&-& \mu_1^2 \left(|\phi_1|^2+|\phi_2|^2\right)
- \mu_3^2 |\phi_3|^2
\nonumber \\ 
&+& \lambda_{11} \left( |\phi_1|^4 
+ |\phi_2|^4 \right)
+ \lambda_{33} |\phi_3|^4 
\nonumber \\ 
&+& \lambda_{1122} |\phi_1|^2 |\phi_2|^2
+ \lambda_{1133}\left(|\phi_1|^2 |\phi_3|^2 + |\phi_2|^2 |\phi_3|^2 \right)
\nonumber \\ 
&+& \lambda_{1221} |\phi_1^{\dagger} \phi_2|^2
+ \lambda_{2332} \left( |\phi_2^{\dagger} \phi_3|^2 + |\phi_1^{\dagger} \phi_3|^2 \right)
\nonumber \\ 
&+& \lambda_{2131} \left( (\phi_2^{\dagger} \phi_1) (\phi_3^{\dagger} \phi_1) - (\phi_1^{\dagger} \phi_2) (\phi_3^{\dagger} \phi_2) \right.
\nonumber \\ 
&+& \left. (\phi_1^{\dagger} \phi_2) (\phi_1^{\dagger} \phi_3) - (\phi_2^{\dagger} \phi_1) (\phi_2^{\dagger} \phi_3) \right)
\nonumber \\ 
&+& \lambda_{1323} (\phi_1^{\dagger} \phi_3) (\phi_2^{\dagger} \phi_3) + \lambda^*_{1323} (\phi_3^{\dagger} \phi_1) (\phi_3^{\dagger} \phi_2),
\end{eqnarray}
where $\lambda_{1323}$ is complex while all other couplings are real.

Another example of a non-Abelian discrete symmetry for the 3HDM
potential is $A_4$, which is a subgroup of~SU(3). This symmetry group
consists of three singlets, $\textbf{1}$, $\textbf{1}^{\prime}$ and
$\textbf{1}^{\prime\prime}$, and one triplet $\textbf{3}$. The
$\textbf{3} \otimes \textbf{3}$ tensor product of $A_4$ decomposes:
\begin{eqnarray}
A_4: \quad \textbf{3} \otimes \textbf{3}=\textbf{1} \oplus \textbf{1}^{\prime} \oplus \textbf{1}^{\prime\prime} \oplus \textbf{3} \oplus\textbf{3}^{\prime}.
\end{eqnarray}
The generators of the $A_4$ discrete symmetry group in terms of double tensor products are
\begin{equation}
\Delta_{A_4}^1=\sigma^0 \otimes {g_1} \otimes \sigma^0, \quad
\Delta_{A_4}^2=\sigma^0 \otimes {g_2} \otimes \sigma^0,
\end{equation}
where
\begin{eqnarray} \small
g_1 = 
	\begin{pmatrix}
	0 & 1 & 0 \\
	0 & 0 & 1\\
	1& 0& 0
	\end{pmatrix}\,,
\quad
g_2 = 
	\begin{pmatrix}
	1 & 0 & 0 \\
	0 & -1 & 0 \\
	0 & 0 & -1
	\end{pmatrix},
\end{eqnarray}
satisfy the conditions: $g_1^3=g_2^2=(g_1 \cdot g_2)^3=1$. 
The $A_4$-symmetric blocks in the bilinear-space can be represented as \vspace{-0.2in}
\begin{flalign}
\begin{minipage}{0.40\textwidth}
\begin{eqnarray} \small
\textbf{1} &:& \;
	\begin{pmatrix}
	\phi_1^{\dagger} \phi_1 +\phi_2^{\dagger} \phi_2+\phi_3^{\dagger} \phi_3 
	\end{pmatrix},
\nonumber \\
 \textbf{1}^{\prime} &:& \;
	\begin{pmatrix}
	\phi_1^{\dagger} \phi_1 + \omega^2 \phi_2^{\dagger} \phi_2 + \omega \phi_3^{\dagger} \phi_3 
	\end{pmatrix},
\nonumber \\
\textbf{1}^{\prime\prime} &:& \;
	\begin{pmatrix}
	\phi_1^{\dagger} \phi_1 + \omega \phi_2^{\dagger} \phi_2 + \omega^2 \phi_3^{\dagger} \phi_3 
	\end{pmatrix},
\nonumber
\end{eqnarray}
\end{minipage}
\begin{minipage}{0.40\textwidth}
\begin{eqnarray}\small
\textbf{3}: \;
	\begin{pmatrix}
	R^1_3 \\ 
	R^2_3 \\ 
	R^3_3 \\ 
	\end{pmatrix},
\quad
\textbf{3}^{\prime}: \;
	\begin{pmatrix}
	R^4_3 \\ 
	-R^5_3 \\ 
	R^6_3 \\
	\end{pmatrix}.
	\nonumber
\end{eqnarray}
\end{minipage}
\end{flalign}
Thus, an $A_4$-invariant 3HDM potential may be written as
\begin{eqnarray}
V_{A_4} =&& -M\,{\bf{1}}
	  +\Lambda_{0}\,{\bf{1^{\mathsf{2}}}}
    +\Lambda_{1}\,{\bf{{1^{\prime}}^\ast}\cdot \bf{1^{\prime\prime}}}
	  +\Lambda_{2}\,{\bf{3^{\mathsf{T}}}\cdot \bf{3}}
	  \nonumber\\
  &&  + \Lambda_{3}\,{\bf{3^{\prime \mathsf{T}}}\cdot \bf{3^{\prime}}}
    + \Lambda_{4}\,{\bf{3^{\prime \mathsf{T}}}\cdot \bf{3}}.
\end{eqnarray}
Equivalently, the  $A_4$-symmetric potential can be rewritten as follows:
\begin{align}
V_{A_4} = &
- \mu_1^2 \left(|\phi_1|^2
+|\phi_2|^2
+ |\phi_3|^2\right)
+\lambda_{11} \left( |\phi_1|^4 
+ |\phi_2|^4 
+ |\phi_3|^4 \right)
 	\nonumber \\ &
+ \lambda_{1122} \left(|\phi_1|^2 |\phi_2|^2 
+ |\phi_1|^2 |\phi_3|^2 
 +|\phi_2|^2 |\phi_3|^2 \right)
 	\nonumber \\ &
+ \lambda_{1221} \left( |\phi_1^{\dagger} \phi_2|^2 
+ |\phi_1^{\dagger} \phi_3|^2 
+ |\phi_2^{\dagger} \phi_3|^2\right)
	\nonumber \\ &
+ {\lambda_{1212} \over 2} \left((\phi_1^{\dagger} \phi_2)^2 +
(\phi_1^{\dagger} \phi_3)^2 +
(\phi_2^{\dagger} \phi_3)^2 \right) 
	\nonumber \\ &
+ {\lambda_{1212}^* \over 2} \left( (\phi_2^{\dagger} \phi_1)^2 +
(\phi_3^{\dagger} \phi_1)^2 +
(\phi_3^{\dagger} \phi_2)^2 \right).
\label{pa4}
\end{align}

In a similar way, the remaining 3HDM potentials that are invariant under non-Abelian discrete symmetries may be obtained. These are presented in Appendix~\ref{D}.

In Tables~\ref{tab1} and~\ref{tab2}, we present all
SU(2)$_L$-preserving accidental symmetries for the 2HDM and the 3HDM
potentials.  The 2HDM potential has a total number of 13 accidental
symmetries~\cite{Battye:2011jj}, of which 6 preserve
U(1)$_Y$~\cite{Ivanov:2007de,Ferreira:2009wh,Ferreira:2010yh} and 7
are custodially symmetric~\cite{Pilaftsis:2011ed}. Given the isomorphism
of the Lie algebras: SO(5) $\sim$ Sp(4), the maximal symmetry group of
the 2HDM in the original $\bm{\Phi}$-field space is
$G^{\bm{\Phi}}_{\text{2HDM}} = [\text{Sp(4)}/\text{Z}_2] \otimes
\text{SU(2)}_L$~\cite{Pilaftsis:2011ed}.

For the case of the 3HDM potential, we find that there exists a total
number of $40$ SU(2)$_L$-preserving accidental symmetries, of which
$18$ preserve U(1)$_Y$ and $22$ are custodially symmetric.  The
maximal symmetry group of the 3HDM potential in the original
$\bm{\Phi}$-field space is
$G^{\bm{\Phi}}_{\text{3HDM}} = [\text{Sp(6)}/\text{Z}_2] \otimes
\text{SU(2)}_L$~\cite{Pilaftsis:2011ed}.
Note that the $40$ accidental symmetries are subgroups of Sp(6).

\section{Conclusions} \label{sec:con}

The $n$HDM potentials may realize a large number of SU(2)$_L$-preserving
accidental symmetries as subgroups of the symplectic group
Sp(2$n)$. We have shown that there are {\em{two}} sets of symmetries:
(i) continuous symmetries and (ii) discrete symmetries (Abelian and
non-Abelian symmetry groups). For the continuous symmetries, we have
offered an algorithmic method that provides the full list of proper,
improper and semi-simple subgroups for any given integer $n$. We have
also included all known discrete symmetries in $n$HDM potentials.
 
Having defined the bi-adjoint representation of the Sp($2n$)
symmetry group, we introduced prime invariants and irreducible
representations in the bilinear field space to construct the scalar
sector of $n$HDM potentials. These quantities have been systematically
used to construct accidentally symmetric $n$HDM potentials by
employing fundamental building blocks that respect the symmetries.
 
Using the method presented in this paper, we have been able to
classify all symmetries and the relations among the theoretical
parameters of the scalar potential for the following: (i) the 2HDM and (ii) the
3HDM. For the 2HDM potential, we recover the maximum number of $13$
accidental symmetries. For the 3HDM potential, we derive {\em{
    for the first time}} the complete list of $40$ accidental
symmetries.

Our approach can be systematically applied to $n$HDM potentials, with
$n>3$, once all possible discrete symmetries have been identified.

\section{Acknowledgements}
\noindent
The work of A.P. and N.D. is supported in part by the
Lancaster-Manchester-Sheffield Consortium\- for Fundamental Physics,
under STFC research grant ST/P000800/1. 
\\The work of N.D. is also supported in part by the Polish National Science Centre
HARMONIA grant under contract UMO- 2015/18/M/ST2/00518 (2016-2020).

\appendix

\section{The 2HDM and The 3HDM Potentials}
\label{V2-3}

In Section \ref{sec:s1}, we have shown that the potential $V_n$ for an $n$HDM can be written down in the quadratic form with the help of $n(2n-1)$-vector $R_{n}^{A}$ as
\begin{equation*}
V_n=-\dfrac{1}{2}M^n_{A} R_n^{A}+\dfrac{1}{4}L^n_{{A}{A'}}R_n^{A}R_n^{{A'}},
\end{equation*}
where $M^n_{A} $ is the mass matrix and $L^n_{{A}{A'}}$ is a quartic coupling matrix.

In the case of 2HDM, the general potential is given by
{\small
\begin{align}
V_{\text{2HDM}}=
& 
- \mu_1^2 ( \phi_1^{\dagger} \phi_1) 
- \mu_2^2 ( \phi_2^{\dagger} \phi_2) 
 - \Big[ m_{12}^2 ( \phi_1^{\dagger} \phi_2)\: +\: {\rm H.c.}\Big]
 \nonumber\\&
 + \lambda_1 ( \phi_1^{\dagger} \phi_1)^2
 + \lambda_2 ( \phi_2^{\dagger} \phi_2)^2
 + \lambda_3 ( \phi_1^{\dagger} \phi_1)( \phi_2^{\dagger} \phi_2)
 + \lambda_4 ( \phi_1^{\dagger} \phi_2)( \phi_2^{\dagger} \phi_1) 
 \nonumber \\&
 + \bigg[\, {1 \over 2} \lambda_5 ( \phi_1^{\dagger} \phi_2)^2
 + \lambda_6 ( \phi_1^{\dagger} \phi_1)( \phi_1^{\dagger} \phi_2)
 + \lambda_7 ( \phi_1^{\dagger} \phi_2)( \phi_2^{\dagger} \phi_2)\: +\:
   {\rm H.c.} \bigg]\;.
\end{align}}
Thus, in the bilinear formalism, the mass $M^2_{A}$ and the quartic
couplings $L^2_{{A}{A'}}$ matrices for the 2HDM potential assume the
following forms~\cite{Pilaftsis:2011ed}:
\begin{equation}
\setlength{\arraycolsep}{2pt}
\renewcommand{\arraystretch}{0.3}
M^2_{A}=(\mu_1^2 + \mu_2^2, \, 2\text{Re}(m_{12}^2),\,
-2\text{Im}(m_{12}^2),\ \mu_1^2 -\mu_2^2, \ 0, \ 0),  
\end{equation}
and
{\small\begin{equation}
\setlength{\arraycolsep}{3pt}
\renewcommand{\arraystretch}{0.6}
L^2_{{A}{A^\prime}}=\begin{pmatrix}
\lambda_1 +\lambda_2 +\lambda_3 & \text{Re} (\lambda_6 +\lambda_7) & -\text{Im} (\lambda_6 +\lambda_7) & \lambda_1 - \lambda_2 & . & . \\
\text{Re} (\lambda_6 + \lambda_7) & \lambda_4 +\text{Re} (\lambda_5) & -\text{Im}(\lambda_5) & \text{Re}(\lambda_6 -\lambda_7) & . & . \\
-\text{Im}(\lambda_6 +\lambda_7) & -\text{Im} (\lambda_5) & \lambda_4 -\text{Re} (\lambda_5) & -\text{Im}(\lambda_6 -\lambda_7) & . & .\\
\lambda_1 -\lambda_2 & \text{Re}(\lambda_6 - \lambda_7) & -\text{Im}(\lambda_6 -\lambda_7) & \lambda_1 +\lambda_2 -\lambda_3 & . & .\\
.& . & . & . & . & . \\
. & . & . & . & . & . 
\end{pmatrix}.
\end{equation}}
Evidently, for a $U(1)_Y$-invariant 2HDM potential, not all the
elements of $M^2_{A}$ and $L^2_{{A}{A^\prime}}$ are non-zero, but only
those for which $A, A^\prime=0,1,2,3$.  

In the case of 3HDM, the general potential has the following form:
{\small
\begin{align}
V_{\text{3HDM}} = 
&
- \mu_1^2 (\phi_1^{\dagger} \phi_1)
- \mu_2^2 (\phi_2^{\dagger} \phi_2)
- \mu_3^2 (\phi_3^{\dagger} \phi_3)
 - \Big[ m_{12}^2 (\phi_1^{\dagger} \phi_2)
+m_{13}^2 (\phi_1^{\dagger} \phi_3)
\nonumber\\ &
+m_{23}^2 (\phi_2^{\dagger} \phi_3)\: +\:
   {\rm H.c.} \Big]
+\lambda_{11} (\phi_1^{\dagger} \phi_1)^2 
+ \lambda_{22} (\phi_2^{\dagger} \phi_2)^2 
+ \lambda_{33} (\phi_3^{\dagger} \phi_3)^2
\nonumber \\ &
+ \lambda_{1122} (\phi_1^{\dagger} \phi_1) (\phi_2^{\dagger} \phi_2) 
+ \lambda_{1133} (\phi_1^{\dagger} \phi_1) (\phi_3^{\dagger} \phi_3) 
+ \lambda_{2233} (\phi_2^{\dagger} \phi_2) (\phi_3^{\dagger} \phi_3) 
\nonumber \\ &
+ \lambda_{1221} (\phi_1^{\dagger} \phi_2) (\phi_2^{\dagger} \phi_1)
+ \lambda_{1331} (\phi_1^{\dagger} \phi_3) (\phi_3^{\dagger} \phi_1) 
+ \lambda_{2332} (\phi_2^{\dagger} \phi_3) (\phi_3^{\dagger} \phi_2)
\nonumber \\ &
 + \bigg[\, {\lambda_{1212} \over 2} (\phi_1^{\dagger} \phi_2)^2 
+ {\lambda_{1313} \over 2} (\phi_1^{\dagger} \phi_3)^2 
+ {\lambda_{2323} \over 2}  (\phi_2^{\dagger} \phi_3)^2 
\nonumber \\ & 
+ \lambda_{1213}  (\phi_1^{\dagger} \phi_2) (\phi_1^{\dagger} \phi_3)
+ \lambda_{2113} (\phi_2^{\dagger} \phi_1) (\phi_1^{\dagger} \phi_3)
+ \lambda_{1323}    (\phi_1^{\dagger} \phi_3) (\phi_2^{\dagger} \phi_3)
\nonumber \\ & 
+ \lambda_{1332}   (\phi_1^{\dagger} \phi_3) (\phi_3^{\dagger} \phi_2)
+ \lambda_{2123}  (\phi_2^{\dagger} \phi_1) (\phi_2^{\dagger} \phi_3)
+ \lambda_{1223} (\phi_1^{\dagger} \phi_2) (\phi_2^{\dagger} \phi_3) 
\nonumber \\ & 
+ \lambda_{1112}  (\phi_1^{\dagger} \phi_1) (\phi_1^{\dagger} \phi_2)
+ \lambda_{2212}  (\phi_2^{\dagger} \phi_2) (\phi_1^{\dagger} \phi_2)  
+ \lambda_{1113}  (\phi_1^{\dagger} \phi_1) (\phi_1^{\dagger} \phi_3) 
\nonumber \\ & 
+ \lambda_{1123}  (\phi_1^{\dagger} \phi_1) (\phi_2^{\dagger} \phi_3)
+ \lambda_{2213}  (\phi_2^{\dagger} \phi_2) (\phi_1^{\dagger} \phi_3)  
+ \lambda_{2223}  (\phi_2^{\dagger} \phi_2) (\phi_2^{\dagger} \phi_3) 
\nonumber \\ & 
+ \lambda_{3312}  (\phi_3^{\dagger} \phi_3) (\phi_1^{\dagger} \phi_2) 
+ \lambda_{3313}  (\phi_3^{\dagger} \phi_3)(\phi_1^{\dagger} \phi_3) 
+ \lambda_{3323}  (\phi_3^{\dagger} \phi_3) (\phi_2^{\dagger} \phi_3) 
+{\rm H.c.}\bigg]\;.
\end{align}}
In the bilinear formalism for this model, the mass $M^3_{A}$ and the
quartic couplings $L^3_{{A}{A'}}$ matrices are given by 
\begin{eqnarray}
\setlength{\arraycolsep}{2pt}
\renewcommand{\arraystretch}{0.3}
M^3_{A} &= ( \mu_1^2 + \mu_2^2+ \mu_3^2, \, 2\text{Re}(m_{12}^2),\, 2\text{Re}(m_{13}^2),\, 2\text{Re}(m_{23}^2),\, -2\text{Im}(m_{12}^2),\, -2\text{Im}(m_{13}^2),\, 
\nonumber \\ 
& -2\text{Im}(m_{23}^2),\, \mu_1^2 -\mu_2^2, \, {1 \over \sqrt{3}} (\mu_1^2 + \mu_2^2-2 \mu_3^2),\, 0 , 0 , 0 , 0 , 0 , 0 ),
\nonumber \\ 
\end{eqnarray}
and 
\begin{align}
&\setlength{\arraycolsep}{3pt}
\renewcommand{\arraystretch}{0.6}
L^3_{{A}{A'}} = \left(
\begin{array}{ccccccccccccccc}
  a_{00} &  a_{01} &  a_{02} &  a_{03} &  a_{04} &  a_{05} &
 a_{06} &  a_{07} &  a_{08} & . & . & . & . & . & . \\
  a_{01} &  a_{11} &  a_{12} &  a_{13} &  a_{14} &  a_{15} &
   a_{16} &  a_{17} &  a_{18} & . & . & . & . & . & . \\
  a_{02} &  a_{12} &  a_{22} &  a_{23} &  a_{24} &  a_{25} &
   a_{26} &  a_{27} &  a_{28} & . & . & . & . & . & . \\
  a_{03} &  a_{13} &  a_{23} &  a_{33} &  a_{34} &  a_{35} &
   a_{36} &  a_{37} &  a_{38} & . & . & . & . & . & . \\
  a_{04} &  a_{14} &  a_{24} &  a_{34} &  a_{44} &  a_{45} &
   a_{46} &  a_{47} &  a_{48} & . & . & . & . & . & . \\
  a_{05} &  a_{15} &  a_{25} &  a_{35} &  a_{45} &  a_{55} &
   a_{56} &  a_{57} &  a_{58} & . & . & . & . & . & . \\
  a_{06} &  a_{16} &  a_{26} &  a_{36} &  a_{46} &  a_{56} &
   a_{66} &  a_{67} &  a_{68} & . & . & . & . & . & . \\
  a_{07} &  a_{17} &  a_{27} &  a_{37} &  a_{47} &  a_{57} &
   a_{67} &  a_{77} &  a_{78} & . & . & . & . & . & . \\
  a_{08} &  a_{18} &  a_{28} &  a_{38} &  a_{48} &  a_{58} &
   a_{68} &  a_{78} &  a_{88} & . & . & . & . & . & . \\
. & . & . & . & . & . & . & . & . & . & . & . & . & . & . \\
 . & . & . & . & . & . & . & . & . & . & . & . & . & . & . \\
. & . & . & . & . & . & . & . & . & . & . & . & . & . & . \\
. & . & . & . & . & . & . & . & . & . & . & . & . & . & . \\
. & . & . & . & . & . & . & . & . & . & . & . & . & . & . \\
. & . & . & . & . & . & . & . & . & . & . & . & . & . & . 
\end{array}
\right),
\end{align}
with $L^3_{{A}{A'}} = L^3_{{A'}{A}}$ and $A,A' = 0,1,\dots,14$. The
non-zero elements of $L^3_{{A}{A'}}$ are:

{\small
\begin{tabular}{l c l}
 & &  \\[0.05mm]
  $a_{00} = \frac{4}{9} ( { \lambda_{11}} + { \lambda_{1122}} + { \lambda_{1133}}+ { \lambda_{22}}+ 
  { \lambda_{2233}}+ { \lambda_{33}})$ 
  & &
  $a_{01} = \frac{2}{3} \text{Re}( { \lambda_{1112}} + { \lambda_{2212}} + { \lambda_{3312}})$ 
  \\[0.1mm]
 $a_{02} = \frac{2}{3}
  \text{Re}( { \lambda_{1113}} + { \lambda_{2213}} + { \lambda_{3313}})$ 
  & &
  $a_{03} = \frac{2}{3}
  \text{Re}( { \lambda_{1123}} + { \lambda_{2223}} + { \lambda_{3323}})$ 
 \\[0.1mm]
  $a_{04} = - \frac{2}{3}
  \text{Im}( { \lambda_{1112}} + { \lambda_{2212}} + { \lambda_{3312}} )$ 
 & &
  $a_{05} = - \frac{2}{3} 
  \text{Im}( { \lambda_{1113}} + { \lambda_{2213}} + { \lambda_{3313}})$ 
 \\[0.1mm]
  $a_{06} = - \frac{2}{3}
  \text{Im}( { \lambda_{1123}} + { \lambda_{2223}} + { \lambda_{3323}})$ 
  & &
  $a_{07} = \frac{1}{3}
  ( 2{ \lambda_{11}} + { \lambda_{1133}}- 2 { \lambda_{22}}- { \lambda_{2233}})$ 
 \\[0.1mm]
  $a_{08} = \frac{\sqrt{3}}{9} \left[ 2{ \lambda_{11}} + 2 
  { \lambda_{1122}} - { \lambda_{1133}} + 2{ \lambda_{22}} - { \lambda_{2233}} - 4{ \lambda_{33}} \right]$
 & &
  $a_{11} =\text{Re}({ \lambda_{1212}}) + { \lambda_{1221}}$ 
 \\[0.1mm]
  $a_{12} = \text{Re}( { \lambda_{1213}} + { \lambda_{2113}} )$ 
  & &
 $a_{13} = \text{Re}( { \lambda_{1223}} + { \lambda_{2123}})$ 
 \\[0.1mm]
 $a_{14} = -\text{Im}(\lambda_{1212})$ 
  & &
 $a_{15} = \text{Im}({ \lambda_{2113}} - { \lambda_{1213}})$ 
 \\[0.1mm]
 $a_{16} = \text{Im}( { \lambda_{2123}} - { \lambda_{1223}})$ 
  & &
 $a_{17} = \text{Re}( { \lambda_{1112}} - { \lambda_{2212}})$ 
 \\[0.1mm]
 $a_{18} = \frac{1}{\sqrt{3}} \text{Re}( 
  { \lambda_{1112}} + { \lambda_{2212}} -2 { \lambda_{3312}})$ 
  & &
 $a_{22} = \text{Re}( { \lambda_{1313}}) + 2 { \lambda_{1331}}$
  \\[0.1mm]
   $a_{23} = 
  2\text{Re}( { \lambda_{1323}} + { \lambda_{1332}})$ 
  & &
 $a_{24} = \text{Im}( { \lambda_{2113}} - { \lambda_{1213}})$ 
  \\[0.1mm]
 $a_{25} =- \text{Im}( { \lambda_{1313}})$ 
  & &
 $a_{26} = \text{Im}( \lambda_{1332} - \lambda_{1323})$  
 \\[0.1mm]
 $a_{27} = \text{Re}( { \lambda_{1113}} - { \lambda_{2213}})$  
  & &
 $a_{28} = \frac{1}{\sqrt{3}} \text{Re}(
  { \lambda_{1113}} + { \lambda_{2213}} -2 { \lambda_{3313}})$ 
 \\[0.1mm] 
 $a_{33} = \text{Re}( { \lambda_{2323}} + 2 { \lambda_{2332}})$  
  & &
 $a_{34} = \text{Im}({ \lambda_{2123}} - \lambda_{1223})$  
 \\[0.1mm]
 $a_{35} = - \text{Im}( { \lambda_{1323}} + { \lambda_{1332}})$  
  & &
 $a_{36} = - \text{Im}( { \lambda_{2323}})$  
 \\[0.1mm]
 $a_{37} = \text{Re}( { \lambda_{1123}} - { \lambda_{2223}})$  
  & &
 $a_{38} = \frac{1}{\sqrt{3}} \text{Re}( 
  { \lambda_{1123}} +\ { \lambda_{2223}} - 2 { \lambda_{3323}} )$ 
 \\[0.1mm]
 $a_{44} = 2 { \lambda_{1221}} - \text{Re}({ \lambda_{1212}})$  
  & &
 $a_{45} = \text{Re}( { \lambda_{2113}}- { \lambda_{1213}})$  
 \\[0.1mm]
 $a_{46} = \text{Re}({ \lambda_{2123}} - { \lambda_{1223}})$  
  & &
 $a_{47} =\text{Im}( { \lambda_{2212}} - { \lambda_{1112}})$  
 \\[0.1mm]
 $a_{48} = \frac{1}{\sqrt{3}} \text{Im}( 2 { \lambda_{3312}} -
  { \lambda_{1112}} - { \lambda_{2212}})$  
  & &
 $a_{55} = { \lambda_{1331}} - \text{Re}({ \lambda_{1313}}) + $  
 \\[0.1mm]
 $a_{56} = \text{Re}({ \lambda_{1332}}- { \lambda_{1323}})$  
  & &
 $a_{57} = \text{Im}({ \lambda_{2213}}- { \lambda_{1113}})$  
 \\[0.1mm]
 $a_{58} = \frac{1}{\sqrt{3}} \text{Im}(2{ \lambda_{3313}}
  -{ \lambda_{1113}}-{ \lambda_{2213}})$  
  & &
 $a_{66} = { \lambda_{2332}} - \text{Re}({ \lambda_{2323}})$  
 \\[0.1mm]
$a_{67} = \text{Im}( { \lambda_{2223}} - { \lambda_{1123}})$  
  & &
 $a_{68} = \frac{1}{\sqrt{3}} \text{Im}(2 { \lambda_{3323}} -
  { \lambda_{1123}} - { \lambda_{2223}})$  
 \\[0.1mm]
 $a_{77} = { \lambda_{11}}+ { \lambda_{22}} -2 { \lambda_{1122}}$  
  & &
 $a_{78} = \frac{1}{\sqrt{3}} ( { \lambda_{11}} - { \lambda_{22}} - 2 { \lambda_{1133}} + 2 { \lambda_{2233}} )$  
 \\[0.1mm]
 $a_{88} = 
  \frac{1}{3} ( { \lambda_{11}} + { \lambda_{22}} + 4 {\lambda_{33}} + 2 { \lambda_{1122}} - 4 { \lambda_{1133}} - 4 { \lambda_{2233}})$
   \\[0.1mm]
\end{tabular}
}
\\

Note that the remaining elements denoted by dots are zero.
Specifically, for a $U(1)_Y$-invariant 3HDM potential, all elements of
$M^3_{A}$ and $L^3_{{A}{A^\prime}}$ corresponding to
$A, A^\prime=9, 10, \dots, 14$ vanish.

\section{The bi-adjoint representations of Sp(4) and Sp(6)}
\label{sp-g}

In Section {\ref{sec:s1}, we introduced the Sp(2$n$) generators in the bi-adjoint representation as
\begin{equation*}
(T^{{B}}_{n})_{IJ}\ =\ -i f_n^{{B}IJ}\ =\ \text{Tr}([ {\Sigma}^I_n, K^{{B}}_n]{\Sigma}^{J}_n). 
\end{equation*}
The maximal symmetry of the potential in the case of 2HDM is Sp(4).
With the help of the above relation, we may derive the following 10
generators in the bi-adjoint representation of Sp(4):
 \begin{align*}
T^0 &=\scriptsize \begin{pmatrix}
0 & 0 & 0 & 0 & 0 \\
0 & 0 & 0 & 0 & 0 \\
0 & 0 & 0 & 0 & 0 \\
0 & 0 & 0 & 0 & i \\
0 & 0 & 0 & -i & 0 
\end{pmatrix}, \qquad 
T^1 =\begin{pmatrix}
0 & 0 & 0 & 0 & 0 \\
0 & 0 & -i & 0 & 0 \\
0 & i & 0 & 0 & 0 \\
0 & 0 & 0 & 0 & 0 \\
0 & 0 & 0 & 0 & 0 \\
\end{pmatrix}, \qquad 
T^2=\begin{pmatrix}
0 & 0 & i & 0 & 0 \\
0 & 0 & 0 & 0 & 0 \\
-i & 0 & 0 & 0 & 0 \\
0 & 0 & 0 & 0 & 0 \\
0 & 0 & 0 & 0 & 0 
\end{pmatrix},
\\[.2mm]
T^3 &=\scriptsize\begin{pmatrix}
0 & -i & 0 & 0 & 0 \\
i & 0 & 0 & 0 & 0 \\
0 & 0 & 0 & 0 & 0 \\
0 & 0 & 0 & 0 & 0 \\
0 & 0 & 0 & 0 & 0 
\end{pmatrix}, \qquad
T^4=\begin{pmatrix}
0 & 0 & 0 & 0 & 0 \\
0 & 0 & 0 & -i & 0 \\
0 & 0 & 0 & 0 & 0 \\
0 & i & 0 & 0 & 0 \\
0 & 0 & 0 & 0 & 0 
\end{pmatrix}, \qquad 
T^5=\begin{pmatrix}
0 & 0 & 0 & 0 & i \\
0 & 0 & 0 & 0 & 0 \\
0 & 0 & 0 & 0 & 0 \\
0 & 0 & 0 & 0 & 0 \\
-i & 0 & 0 & 0 & 0 
\end{pmatrix},\\[.2mm]
T^6 &=\scriptsize\begin{pmatrix}
0 & 0 & 0 & 0 & 0 \\
0 & 0 & 0 & 0 & i \\
0 & 0 & 0 & 0 & 0 \\
0 & 0 & 0 & 0 & 0 \\
0 & -i & 0 & 0 & 0 
\end{pmatrix}, \qquad 
T^7 =\begin{pmatrix}
0 & 0 & 0 & i & 0 \\
0 & 0 & 0 & 0 & 0 \\
0 & 0 & 0 & 0 & 0 \\
-i & 0 & 0 & 0 & 0 \\
0 & 0 & 0 & 0 & 0 
\end{pmatrix}, \qquad
T^8 =\begin{pmatrix}
0 & 0 & 0 & 0 & 0 \\
0 & 0 & 0 & 0 & 0 \\
0 & 0 & 0 & 0 & -i \\
0 & 0 & 0 & 0 & 0 \\
0 & 0 & i & 0 & 0 
\end{pmatrix}, \\[.2mm]
T^9 &=\scriptsize\begin{pmatrix}
0 & 0 & 0 & 0 & 0 \\
0 & 0 & 0 & 0 & 0 \\
0 & 0 & 0 & -i & 0 \\
0 & 0 & i & 0 & 0 \\
0 & 0 & 0 & 0 & 0
\end{pmatrix}. 
\end{align*}
Note that these generators are identical to those of the SO(5) group
in the fundamental representation~\cite{Pilaftsis:2011ed}, thereby establishing the local
group isomorphism: Sp(4) $\cong$ SO(5).

For the case of the 3HDM, the maximal symmetry is Sp(6). Similarly, 
the 21 generators of Sp(6) in the bi-adjoint representation read: 
{\small \begin{align*}
&\setlength{\arraycolsep}{1.5pt}
\renewcommand{\arraystretch}{0.3}
T^{0}=i\left(\scriptsize
\begin{array}{cccccccccccccc}
 . &. & . & . & . & . & . & . & . & . & . & . & . & . \\
. &. & . & . & . & . & . & . & . & . & . & . & . & .\\
. &. & . & . & . & . & . & . & . & . & . & . & . & . \\
. &. & . & . & . & . & . & . & . & . & . & . & . & . \\
. &. & . & . & . & . & . & . & . & . & . & . & . & . \\
. &. & . & . & . & . & . & . & . & . & . & . & . & . \\
. &. & . & . & . & . & . & . & . & . & . & . & . & . \\
. &. & . & . & . & . & . & . & . & . & . & . & . & . \\
. & . &. & . &. & . & . & . & . & . & . &\textbf{ 2} & . & . \\
 . & . & . & . & . & . & . & . & . & . & . & . &\textbf{ 2} &. \\
 . & . & . & . & . & . & . & . & . & . & . & . & . &\textbf{ 2 }\\
 . & . & . & . & . & . & . & . & \textbf{-2 }& . & . & . & . & . \\
 . & . & . & . & . & . & . & . & . & \textbf{-2} & . & . & . & . \\
 . & . & . & . & . & . & . & . & . & . & \textbf{-2} & . & . & . \\
\end{array}
\right), \quad &&\setlength{\arraycolsep}{1.5pt}
  \renewcommand{\arraystretch}{0.3}
  T^{1}=i\left(\scriptsize
\begin{array}{cccccccccccccc}
 . & . & . & . & . & . & . & . & . & . & . & . & . & . \\
 . & . & . & . & . &\textbf{ -1} & . & . & . & . & . & . & . & . \\
 . & . & . & . & \textbf{-1} & . & . & . & . & . & . & . & . & . \\
 . & . & . & . & . & . & \textbf{-2} & . & . & . & . & . & . & . \\
 . & . & \textbf{1} & . & . & . & . & . & . & . & . & . & . & . \\
. &\textbf{ 1} & . & . & . & . & . & . & . & . & . & . & . & . \\
 . & . & . & \textbf{2} & . & . & . & . & . & . & . & . & . & . \\
 . & . & . & . & . & . & . & . & . & . & . & . & . & . \\
 . & . & . & . & . & . & . & . & . & . & . & . & . & . \\
 . & . & . & . & . & . & . & . & . & . & . & . & . &\textbf{ 1} \\
 . & . & . & . & . & . & . & . & . & . & . & . &\textbf{ 1}& . \\
 . & . & . & . & . & . & . & . & . & . & . & . & . & . \\
 . & . & . & . & . & . & . & . & . & . & \textbf{-1 }& . & . & . \\
 . & . & . & . & . & . & . & . & . & \textbf{-1} & . & . & . & . \\
\end{array}
\right),
\\[0.1cm]
&\setlength{\arraycolsep}{1.5pt}
  \renewcommand{\arraystretch}{0.43}
T^{2}=i\left(\scriptsize
\begin{array}{cccccccccccccc}
 . & . & . & \textbf{-2 }& . & . & . & . & . & . & . & . & . & . \\
 . & . & . & . &\textbf{ -1} & . & . & . & . & . & . & . & . & . \\
 . & . & . & . & . &\textbf{ 1}& . & . & . & . & . & . & . & . \\
 \textbf{2} & . & . & . & . & . & . & . & . & . & . & . & . & . \\
. &\textbf{ 1} & . & . & . & . & . & . & . & . & . & . & . & . \\
 . & . & \textbf{-1} & . & . & . & . & . & . & . & . & . & . & . \\
 . & . & . & . & . & . & . & . & . & . & . & . & . & . \\
 . & . & . & . & . & . & . & . & . & . & . & . & . & . \\
 . & . & . & . & . & . & . & . & . & . & . & . & . & . \\
 . & . & . & . & . & . & . & . & . & . & . & . &\textbf{ 1}& . \\
 . & . & . & . & . & . & . & . & . & . & . & . & . & \textbf{-1} \\
 . & . & . & . & . & . & . & . & . & . & . & . & . & . \\
 . & . & . & . & . & . & . & . & . & \textbf{-1}& . & . & . & . \\
 . & . & . & . & . & . & . & . & . & . &\textbf{ 1}& . & . & . \\
\end{array}
\right),\quad &&\setlength{\arraycolsep}{1.5pt}
  \renewcommand{\arraystretch}{0.3} T^{3}=i\left(\scriptsize
\begin{array}{cccccccccccccc}
 . & . & . & . & . &\textbf{1}& . & . & . & . & . & . & . & . \\
 . & . & . & . & . & . & . & . & . & . & . & . & . & . \\
 . & . & . & \textbf{-1}& . & . & . & . & . & . & . & . & . & . \\
 . & . &\textbf{1}& . & . & . & . & . & . & . & . & . & . & . \\
 . & . & . & . & . & . & \textbf{-1}& \textbf{-$\sqrt{3}$ }& . & . & . & . & . & . \\
 \textbf{-1}& . & . & . & . & . & . & . & . & . & . & . & . & . \\
 . & . & . & . &\textbf{1}& . & . & . & . & . & . & . & . & . \\
 . & . & . & . & \textbf{$\sqrt{3}$ }& . & . & . & . & . & . & . & . & . \\
 . & . & . & . & . & . & . & . & . & . & . & . & . & \textbf{-1} \\
 . & . & . & . & . & . & . & . & . & . & . & . & . & . \\
 . & . & . & . & . & . & . & . & . & . & . & \textbf{-1}& . & . \\
 . & . & . & . & . & . & . & . & . & . &\textbf{1}& . & . & . \\
 . & . & . & . & . & . & . & . & . & . & . & . & . & . \\
 . & . & . & . & . & . & . & . &\textbf{1}& . & . & . & . & . \\
\end{array}
\right),
\\[0.1cm]
&\setlength{\arraycolsep}{1.5pt}
  \renewcommand{\arraystretch}{0.3}
T^{4}=i\left(\scriptsize
\begin{array}{cccccccccccccc}
 . & . & . & . &\textbf{1}& . & . & . & . & . & . & . & . & . \\
 . & . & . &\textbf{1}& . & . & . & . & . & . & . & . & . & . \\
 . & . & . & . & . & . & . & . & . & . & . & . & . & . \\
 . & \textbf{-1}& . & . & . & . & . & . & . & . & . & . & . & . \\
 \textbf{-1}& . & . & . & . & . & . & . & . & . & . & . & . & . \\
 . & . & . & . & . & . &\textbf{1}& \textbf{-$\sqrt{3}$ }& . & . & . & . & . & . \\
 . & . & . & . & . & \textbf{-1}& . & . & . & . & . & . & . & . \\
 . & . & . & . & . & \textbf{$\sqrt{3}$ }& . & . & . & . & . & . & . & . \\
 . & . & . & . & . & . & . & . & . & . & . & . &\textbf{1}& . \\
 . & . & . & . & . & . & . & . & . & . & . &\textbf{1}& . & . \\
 . & . & . & . & . & . & . & . & . & . & . & . & . & . \\
 . & . & . & . & . & . & . & . & . & \textbf{-1}& . & . & . & . \\
 . & . & . & . & . & . & . & . & \textbf{-1}& . & . & . & . & . \\
 . & . & . & . & . & . & . & . & . & . & . & . & . & . \\
\end{array}
\right),\quad &&
\setlength{\arraycolsep}{1.5pt}
  \renewcommand{\arraystretch}{0.3}
T^{5}=\scriptsize {\frac{i}{\sqrt{3}}}\left(\scriptsize
\begin{array}{cccccccccccccc}
 . & . & . & . & . & . & . & . & . & . & . & . & . & . \\
 . & . & . & . & \textbf{-3}& . & . & . & . & . & . & . & . & . \\
 . & . & . & . & . & \textbf{-3}& . & . & . & . & . & . & . & . \\
 . & . & . & . & . & . & . & . & . & . & . & . & . & . \\
 . & \textbf{3 }& . & . & . & . & . & . & . & . & . & . & . & . \\
 . & . & \textbf{3 }& . & . & . & . & . & . & . & . & . & . & . \\
 . & . & . & . & . & . & . & . & . & . & . & . & . & . \\
 . & . & . & . & . & . & . & . & . & . & . & . & . & . \\
 . & . & . & . & . & . & . & . & . & . & . & \textbf{2} & . & . \\
 . & . & . & . & . & . & . & . & . & . & . & . & -\textbf{1} & . \\
 . & . & . & . & . & . & . & . & . & . & . & . & . & -\textbf{1} \\
 . & . & . & . & . & . & . & . & -\textbf{2} & . & . & . & . & . \\
 . & . & . & . & . & . & . & . & . & \textbf{1} & . & . & . & . \\
 . & . & . & . & . & . & . & . & . & . & \textbf{1} & . & . & . \\
\end{array}
\right),
\end{align*}
\begin{align*}
&\setlength{\arraycolsep}{2.2pt}
  \renewcommand{\arraystretch}{0.3}
T^{6}=i\left(\scriptsize
\begin{array}{cccccccccccccc}
 . & . & . & . & . & . & \textbf{2} & . & . & . & . & . & . & . \\
 . & . & \textbf{-1}& . & . & . & . & . & . & . & . & . & . & . \\
. &\textbf{1}& . & . & . & . & . & . & . & . & . & . & . & . \\
 . & . & . & . & . & . & . & . & . & . & . & . & . & . \\
 . & . & . & . & . & \textbf{-1}& . & . & . & . & . & . & . & . \\
 . & . & . & . &\textbf{1}& . & . & . & . & . & . & . & . & . \\
 \textbf{-2}& . & . & . & . & . & . & . & . & . & . & . & . & . \\
 . & . & . & . & . & . & . & . & . & . & . & . & . & . \\
 . & . & . & . & . & . & . & . & . & . & . & . & . & . \\
 . & . & . & . & . & . & . & . & . & . & \textbf{-1}& . & . & . \\
 . & . & . & . & . & . & . & . & . &\textbf{1}& . & . & . & . \\
 . & . & . & . & . & . & . & . & . & . & . & . & . & . \\
 . & . & . & . & . & . & . & . & . & . & . & . & . & \textbf{-1} \\
 . & . & . & . & . & . & . & . & . & . & . & . &\textbf{1}& . \\
\end{array}
\right), \quad &&
\setlength{\arraycolsep}{1.5pt}
  \renewcommand{\arraystretch}{0.3}
T^{7}=i\left(\scriptsize
\begin{array}{cccccccccccccc}
 . & . & \textbf{-1}& . & . & . & . & . & . & . & . & . & . & . \\
 . & . & . & . & . & . &\textbf{1}& \textbf{$\sqrt{3}$ }& . & . & . & . & . & . \\
 \textbf{1} & . & . & . & . & . & . & . & . & . & . & . & . & . \\
 . & . & . & . & . &\textbf{1}& . & . & . & . & . & . & . & . \\
 . & . & . & . & . & . & . & . & . & . & . & . & . & . \\
 . & . & . & \textbf{-1}& . & . & . & . & . & . & . & . & . & . \\
 . & \textbf{-1}& . & . & . & . & . & . & . & . & . & . & . & . \\
 . & \textbf{-$\sqrt{3}$ }& . & . & . & . & . & . & . & . & . & . & . & . \\
 . & . & . & . & . & . & . & . & . & . &\textbf{1}& . & . & . \\
 . & . & . & . & . & . & . & . & . & . & . & . & . & . \\
 . & . & . & . & . & . & . & . & \textbf{-1}& . & . & . & . & . \\
 . & . & . & . & . & . & . & . & . & . & . & . & . & \textbf{1} \\
 . & . & . & . & . & . & . & . & . & . & . & . & . & . \\
 . & . & . & . & . & . & . & . & . & . & . & \textbf{-1}& . & . \\
\end{array}
\right),
\\[0.1cm]
&\setlength{\arraycolsep}{1.5pt}
  \renewcommand{\arraystretch}{0.3}
T^{8}=i\left(\scriptsize
\begin{array}{cccccccccccccc}
 . & \textbf{-1}& . & . & . & . & . & . & . & . & . & . & . & . \\
 \textbf{1} & . & . & . & . & . & . & . & . & . & . & . & . & . \\
 . & . & . & . & . & . & \textbf{-1}& \textbf{$\sqrt{3}$ }& . & . & . & . & . & . \\
 . & . & . & . & \textbf{-1}& . & . & . & . & . & . & . & . & . \\
 . & . & . &\textbf{1}& . & . & . & . & . & . & . & . & . & . \\
 . & . & . & . & . & . & . & . & . & . & . & . & . & . \\
 . & . &\textbf{1}& . & . & . & . & . & . & . & . & . & . & . \\
 . & . & \textbf{-$\sqrt{3}$ }& . & . & . & . & . & . & . & . & . & . & . \\
 . & . & . & . & . & . & . & . & . & \textbf{-1}& . & . & . & . \\
 . & . & . & . & . & . & . & . &\textbf{1}& . & . & . & . & . \\
 . & . & . & . & . & . & . & . & . & . & . & . & . & . \\
 . & . & . & . & . & . & . & . & . & . & . & . & \textbf{-1}& . \\
 . & . & . & . & . & . & . & . & . & . & . &\textbf{1}& . & . \\
 . & . & . & . & . & . & . & . & . & . & . & . & . & . \\
\end{array}
\right), \quad &&
\setlength{\arraycolsep}{2.8pt}
  \renewcommand{\arraystretch}{0.3}
T^{9}=i\left(\scriptsize
\begin{array}{cccccccccccccc}
 . & . & . & . & . & . & . & . & . & . & . & . & . & . \\
 . & . & . & . & . & . & . & . & . & . & . & . & . & . \\
 . & . & . & . & . & . & . & . & . & . & . & . & . & . \\
 . & . & . & . & . & . & . & . & \textbf{2} & . & . & . & . & . \\
 . & . & . & . & . & . & . & . & . & \textbf{2} & . & . & . & . \\
 . & . & . & . & . & . & . & . & . & . & \textbf{2} & . & . & . \\
 . & . & . & . & . & . & . & . & . & . & . & . & . & . \\
 . & . & . & . & . & . & . & . & . & . & . & . & . & . \\
 . & . & . & \textbf{-2}& . & . & . & . & . & . & . & . & . & . \\
 . & . & . & . & \textbf{-2}& . & . & . & . & . & . & . & . & . \\
 . & . & . & . & . & \textbf{-2}& . & . & . & . & . & . & . & . \\
 . & . & . & . & . & . & . & . & . & . & . & . & . & . \\
 . & . & . & . & . & . & . & . & . & . & . & . & . & . \\
 . & . & . & . & . & . & . & . & . & . & . & . & . & . \\
\end{array}
\right),
\\[0.1cm]
&\setlength{\arraycolsep}{2pt}
  \renewcommand{\arraystretch}{0.3}
T^{10}=i\left(\scriptsize
\begin{array}{cccccccccccccc}
 . & . & . & . & . & . & . & . & . & . & . & . & . & . \\
 . & . & . & . & . & . & . & . & . & . & . & . & . & \textbf{-1} \\
 . & . & . & . & . & . & . & . & . & . & . & . & \textbf{-1}& . \\
 . & . & . & . & . & . & . & . & . & . & . & . & . & . \\
 . & . & . & . & . & . & . & . & . & . &\textbf{1}& . & . & . \\
 . & . & . & . & . & . & . & . & . &\textbf{1}& . & . & . & . \\
 . & . & . & . & . & . & . & . & . & . & . & \textbf{2} & . & . \\
 . & . & . & . & . & . & . & . & . & . & . & . & . & . \\
 . & . & . & . & . & . & . & . & . & . & . & . & . & . \\
 . & . & . & . & . & \textbf{-1}& . & . & . & . & . & . & . & . \\
 . & . & . & . & \textbf{-1}& . & . & . & . & . & . & . & . & . \\
 . & . & . & . & . & . & \textbf{-2}& . & . & . & . & . & . & . \\
 . & . &\textbf{1}& . & . & . & . & . & . & . & . & . & . & . \\
.&\textbf{1}& . & . & . & . & . & . & . & . & . & . & . & . \\
\end{array}
\right), \quad &&
\setlength{\arraycolsep}{2pt}
  \renewcommand{\arraystretch}{0.3}
T^{11}=i\left(\scriptsize
\begin{array}{cccccccccccccc}
 . & . & . & . & . & . & . & . & . & . & . & \textbf{-2}& . & . \\
 . & . & . & . & . & . & . & . & . & . & . & . & \textbf{-1}& . \\
 . & . & . & . & . & . & . & . & . & . & . & . & . & \textbf{1} \\
 . & . & . & . & . & . & . & . & . & . & . & . & . & . \\
 . & . & . & . & . & . & . & . & . &\textbf{1}& . & . & . & . \\
 . & . & . & . & . & . & . & . & . & . & \textbf{-1}& . & . & . \\
 . & . & . & . & . & . & . & . & . & . & . & . & . & . \\
 . & . & . & . & . & . & . & . & . & . & . & . & . & . \\
 . & . & . & . & . & . & . & . & . & . & . & . & . & . \\
 . & . & . & . & \textbf{-1}& . & . & . & . & . & . & . & . & . \\
 . & . & . & . & . &\textbf{1}& . & . & . & . & . & . & . & . \\
\textbf{ 2 }& . & . & . & . & . & . & . & . & . & . & . & . & . \\
 . &\textbf{1}& . & . & . & . & . & . & . & . & . & . & . & . \\
 . & . & \textbf{-1}& . & . & . & . & . & . & . & . & . & . & . \\
\end{array}
\right),
\\[0.1cm]
&\setlength{\arraycolsep}{1.5pt}
  \renewcommand{\arraystretch}{0.3}
T^{12}=i\left(\scriptsize
\begin{array}{cccccccccccccc}
 . & . & . & . & . & . & . & . & . & . & . & . & . & \textbf{1} \\
 . & . & . & . & . & . & . & . & . & . & . & . & . & . \\
 . & . & . & . & . & . & . & . & . & . & . & \textbf{-1}& . & . \\
 . & . & . & . & . & . & . & . & . & . & \textbf{-1}& . & . & . \\
 . & . & . & . & . & . & . & . & . & . & . & . & . & . \\
 . & . & . & . & . & . & . & . & \textbf{-1}& . & . & . & . & . \\
 . & . & . & . & . & . & . & . & . & . & . & . &\textbf{1}& . \\
 . & . & . & . & . & . & . & . & . & . & . & . & \textbf{$\sqrt{3}$ }& . \\
 . & . & . & . & . &\textbf{1}& . & . & . & . & . & . & . & . \\
 . & . & . & . & . & . & . & . & . & . & . & . & . & . \\
 . & . & . &\textbf{1}& . & . & . & . & . & . & . & . & . & . \\
 . & . &\textbf{1}& . & . & . & . & . & . & . & . & . & . & . \\
 . & . & . & . & . & . & \textbf{-1}& \textbf{-$\sqrt{3}$ }& . & . & . & . & . & . \\
 \textbf{-1}& . & . & . & . & . & . & . & . & . & . & . & . & . \\
\end{array}
\right), \quad &&
\setlength{\arraycolsep}{1.5pt}
  \renewcommand{\arraystretch}{0.3}
T^{13}=i\left(\scriptsize
\begin{array}{cccccccccccccc}
 . & . & . & . & . & . & . & . & . & . & . & . &\textbf{1}& . \\
 . & . & . & . & . & . & . & . & . & . & . &\textbf{1}& . & . \\
 . & . & . & . & . & . & . & . & . & . & . & . & . & . \\
 . & . & . & . & . & . & . & . & . &\textbf{1}& . & . & . & . \\
 . & . & . & . & . & . & . & . &\textbf{1}& . & . & . & . & . \\
 . & . & . & . & . & . & . & . & . & . & . & . & . & . \\
 . & . & . & . & . & . & . & . & . & . & . & . & . & \textbf{-1} \\
 . & . & . & . & . & . & . & . & . & . & . & . & . & \textbf{$\sqrt{3}$ }\\
 . & . & . & . & \textbf{-1}& . & . & . & . & . & . & . & . & . \\
 . & . & . & \textbf{-1}& . & . & . & . & . & . & . & . & . & . \\
 . & . & . & . & . & . & . & . & . & . & . & . & . & . \\
 . & \textbf{-1}& . & . & . & . & . & . & . & . & . & . & . & . \\
 \textbf{-1}& . & . & . & . & . & . & . & . & . & . & . & . & . \\
 . & . & . & . & . & . &\textbf{1}& \textbf{-$\sqrt{3}$ }& . & . & . & . & . & . \\
\end{array}
\right),
\\[0.1cm]
& \setlength{\arraycolsep}{1.5pt}
  \renewcommand{\arraystretch}{0.3}
T^{14}=\scriptsize {\frac{i}{\sqrt{3}}}\left(\scriptsize
\begin{array}{cccccccccccccc}
 . & . & . & . & . & . & . & . & . & . & . & . & . & . \\
 . & . & . & . & . & . & . & . & . & . & . & . & \textbf{-3}& . \\
 . & . & . & . & . & . & . & . & . & . & . & . & . & \textbf{-3 }\\
 . & . & . & . & . & . & . & . & \textbf{2} & . & . & . & . & . \\
 . & . & . & . & . & . & . & . & . & -\textbf{1} & . & . & . & . \\
 . & . & . & . & . & . & . & . & . & . & -\textbf{1} & . & . & . \\
 . & . & . & . & . & . & . & . & . & . & . & . & . & . \\
 . & . & . & . & . & . & . & . & . & . & . & . & . & . \\
 . & . & . & -\textbf{2} & . & . & . & . & . & . & . & . & . & . \\
 . & . & . & . & \textbf{1} & . & . & . & . & . & . & . & . & . \\
 . & . & . & . & . & \textbf{1} & . & . & . & . & . & . & . & . \\
 . & . & . & . & . & . & . & . & . & . & . & . & . & . \\
 . & \textbf{3 }& . & . & . & . & . & . & . & . & . & . & . & . \\
 . & . & \textbf{3 }& . & . & . & . & . & . & . & . & . & . & . \\
\end{array}
\right), \quad &&
\setlength{\arraycolsep}{2.8pt}
  \renewcommand{\arraystretch}{0.3}
T^{15}=i\left(\scriptsize
\begin{array}{cccccccccccccc}
 . & . & . & . & . & . & . & . & . & . & . & . & . & . \\
 . & . & . & . & . & . & . & . & . & . & . & . & . & . \\
 . & . & . & . & . & . & . & . & . & . & . & . & . & . \\
 . & . & . & . & . & . & . & . & . & . & . & \textbf{-2}& . & . \\
 . & . & . & . & . & . & . & . & . & . & . & . & \textbf{-2}& . \\
 . & . & . & . & . & . & . & . & . & . & . & . & . & \textbf{-2} \\
 . & . & . & . & . & . & . & . & . & . & . & . & . & . \\
 . & . & . & . & . & . & . & . & . & . & . & . & . & . \\
 . & . & . & . & . & . & . & . & . & . & . & . & . & . \\
 . & . & . & . & . & . & . & . & . & . & . & . & . & . \\
 . & . & . & . & . & . & . & . & . & . & . & . & . & . \\
 . & . & . & \textbf{2} & . & . & . & . & . & . & . & . & . & . \\
 . & . & . & . & \textbf{2} & . & . & . & . & . & . & . & . & . \\
 . & . & . & . & . & \textbf{2} & . & . & . & . & . & . & . & . \\
\end{array}
\right),
\\[0.1cm]
& \setlength{\arraycolsep}{2.5pt}
  \renewcommand{\arraystretch}{0.3}
T^{16}=i\left(\scriptsize
\begin{array}{cccccccccccccc}
 . & . & . & . & . & . & . & . & . & . & . & . & . & . \\
 . & . & . & . & . & . & . & . & . & . & \textbf{-1}& . & . & . \\
 . & . & . & . & . & . & . & . & . & \textbf{-1}& . & . & . & . \\
 . & . & . & . & . & . & . & . & . & . & . & . & . & . \\
 . & . & . & . & . & . & . & . & . & . & . & . & . & \textbf{-1} \\
 . & . & . & . & . & . & . & . & . & . & . & . & \textbf{-1}& . \\
 . & . & . & . & . & . & . & . & \textbf{2} & . & . & . & . & . \\
 . & . & . & . & . & . & . & . & . & . & . & . & . & . \\
 . & . & . & . & . & . & \textbf{-2}& . & . & . & . & . & . & . \\
 . & . &\textbf{1}& . & . & . & . & . & . & . & . & . & . & . \\
. &\textbf{1}& . & . & . & . & . & . & . & . & . & . & . & . \\
 . & . & . & . & . & . & . & . & . & . & . & . & . & . \\
 . & . & . & . & . &\textbf{1}& . & . & . & . & . & . & . & . \\
 . & . & . & . &\textbf{1}& . & . & . & . & . & . & . & . & . \\
\end{array}
\right), \quad &&
\setlength{\arraycolsep}{2pt}
  \renewcommand{\arraystretch}{0.3}
T^{17}=i\left(\scriptsize
\begin{array}{cccccccccccccc}
 . & . & . & . & . & . & . & . & \textbf{-2}& . & . & . & . & . \\
 . & . & . & . & . & . & . & . & . & \textbf{-1}& . & . & . & . \\
 . & . & . & . & . & . & . & . & . & . &\textbf{1}& . & . & . \\
 . & . & . & . & . & . & . & . & . & . & . & . & . & . \\
 . & . & . & . & . & . & . & . & . & . & . & . & \textbf{-1}& . \\
 . & . & . & . & . & . & . & . & . & . & . & . & . & \textbf{1} \\
 . & . & . & . & . & . & . & . & . & . & . & . & . & . \\
 . & . & . & . & . & . & . & . & . & . & . & . & . & . \\
\textbf{ 2} & . & . & . & . & . & . & . & . & . & . & . & . & . \\
. &\textbf{1}& . & . & . & . & . & . & . & . & . & . & . & . \\
 . & . & \textbf{-1}& . & . & . & . & . & . & . & . & . & . & . \\
 . & . & . & . & . & . & . & . & . & . & . & . & . & . \\
 . & . & . & . &\textbf{1}& . & . & . & . & . & . & . & . & . \\
 . & . & . & . & . & \textbf{-1}& . & . & . & . & . & . & . & . \\
\end{array}
\right),
\\[0.1cm]
& \setlength{\arraycolsep}{1.5pt}
  \renewcommand{\arraystretch}{0.3}
T^{18}=i\left(\scriptsize
\begin{array}{cccccccccccccc}
 . & . & . & . & . & . & . & . & . & . &\textbf{1}& . & . & . \\
 . & . & . & . & . & . & . & . & . & . & . & . & . & . \\
 . & . & . & . & . & . & . & . & \textbf{-1}& . & . & . & . & . \\
 . & . & . & . & . & . & . & . & . & . & . & . & . & \textbf{1} \\
 . & . & . & . & . & . & . & . & . & . & . & . & . & . \\
 . & . & . & . & . & . & . & . & . & . & . &\textbf{1}& . & . \\
 . & . & . & . & . & . & . & . & . &\textbf{1}& . & . & . & . \\
 . & . & . & . & . & . & . & . & . & \textbf{$\sqrt{3}$ }& . & . & . & . \\
 . & . &\textbf{1}& . & . & . & . & . & . & . & . & . & . & . \\
 . & . & . & . & . & . & \textbf{-1}& \textbf{-$\sqrt{3}$ }& . & . & . & . & . & . \\
 \textbf{-1}& . & . & . & . & . & . & . & . & . & . & . & . & . \\
 . & . & . & . & . & \textbf{-1}& . & . & . & . & . & . & . & . \\
 . & . & . & . & . & . & . & . & . & . & . & . & . & . \\
 . & . & . & \textbf{-1}& . & . & . & . & . & . & . & . & . & . \\
\end{array}
\right), \quad &&
\setlength{\arraycolsep}{1.5pt}
  \renewcommand{\arraystretch}{0.3}
T^{19}=i\left(\scriptsize
\begin{array}{cccccccccccccc}
 . & . & . & . & . & . & . & . & . &\textbf{1}& . & . & . & . \\
 . & . & . & . & . & . & . & . &\textbf{1}& . & . & . & . & . \\
 . & . & . & . & . & . & . & . & . & . & . & . & . & . \\
 . & . & . & . & . & . & . & . & . & . & . & . & \textbf{-1}& . \\
 . & . & . & . & . & . & . & . & . & . & . & \textbf{-1}& . & . \\
 . & . & . & . & . & . & . & . & . & . & . & . & . & . \\
 . & . & . & . & . & . & . & . & . & . & \textbf{-1}& . & . & . \\
 . & . & . & . & . & . & . & . & . & . & \textbf{$\sqrt{3}$ }& . & . & . \\
 . & \textbf{-1}& . & . & . & . & . & . & . & . & . & . & . & . \\
 \textbf{-1}& . & . & . & . & . & . & . & . & . & . & . & . & . \\
 . & . & . & . & . & . &\textbf{1}& \textbf{-$\sqrt{3}$ }& . & . & . & . & . & . \\
 . & . & . & . &\textbf{1}& . & . & . & . & . & . & . & . & . \\
 . & . & . &\textbf{1}& . & . & . & . & . & . & . & . & . & . \\
 . & . & . & . & . & . & . & . & . & . & . & . & . & . \\
\end{array}
\right),
\end{align*}
\begin{align*}
& \setlength{\arraycolsep}{1.5pt}
  \renewcommand{\arraystretch}{0.3}
T^{20}=\scriptsize {\frac{i}{\sqrt{3}}}\left(\scriptsize
\begin{array}{cccccccccccccc}
 . & . & . & . & . & . & . & . & . & . & . & . & . & . \\
 . & . & . & . & . & . & . & . & . & \textbf{- 3 }& . & . & . & . \\
 . & . & . & . & . & . & . & . & . & . & \textbf{-3 }& . & . & . \\
 . & . & . & . & . & . & . & . & . & . & . & -\textbf{2} & . & . \\
 . & . & . & . & . & . & . & . & . & . & . & . & \textbf{1}& . \\
 . & . & . & . & . & . & . & . & . & . & . & . & . & \textbf{1} \\
 . & . & . & . & . & . & . & . & . & . & . & . & . & . \\
 . & . & . & . & . & . & . & . & . & . & . & . & . & . \\
 . & . & . & . & . & . & . & . & . & . & . & . & . & . \\
 . & \textbf{3 }& . & . & . & . & . & . & . & . & . & . & . & . \\
 . & . & \textbf{3 } & . & . & . & . & . & . & . & . & . & . & . \\
 . & . & . &\textbf{2 }& . & . & . & . & . & . & . & . & . & . \\
 . & . & . & . & -\textbf{1} & . & . & . & . & . & . & . & . & . \\
 . & . & . & . & . & -\textbf{1}& . & . & . & . & . & . & . & . \\
\end{array}
\right).
\end{align*}}

\section{Prime invariants and symmetric $n$HDM potentials}\label{C}

In Section \ref{sec:cs}, we have shown that one can systematically
construct symmetric $n$HDM potentials in terms of fundamental building
blocks.

Here, we will employ maximal and minimal invariants to construct 
2HDM potentials invariant under continuous symmetries. The symmetry
groups and the respective functional forms of the 2HDM potentials in
terms of prime invariants may be listed as follows:
\begin{itemize}
\item
CP1 $\otimes$ SO(2):
$V[S_{11}+S_{22},D_{12}^2, T^2_{12}$].
\item
SU(2):
$V[S_{11}+S_{22},D_{12}^2]$.
\item
Sp(2$)_{\phi_1 +\phi_2} \subset$ Sp(2$)_{\phi_1} \otimes$ Sp(2$)_{\phi_2}$: $V[S_{11},S_{22},S_{12}]$.
\item
U(1$)_{\text{}} \otimes$ Sp(2)$_{\phi_1 \phi_2}$: $V[S_{11}+S_{22},D_{12}^{\prime 2}]$.
\item
Sp(2$)_{\phi_1} \otimes$ Sp(2$)_{\phi_2}$: $V[S_{11},S_{22}]$.
\item
 Sp(4): $V[S_{11}+S_{22}]$.
\end{itemize}

\begin{table}[t]
\small
\centering
\begin{tabular}{| c | c c | c l |}
\hline
No. & Symmetry & & & Non-zero parameters for 2HDM potentials\\
\hline \hline
1 & CP1 & & & $\mu_1^2$, $\mu_2^2$, Re($m_{12}^2$), $\lambda_1$, $\lambda_2$, $\lambda_3$, $\lambda_4$, Re($\lambda_5$), Re($\lambda_6$), Re($\lambda_7$) \\ \hline
2 & $Z_2$ & & & $\mu_1^2$, $\mu_2^2$, $\lambda_1$, $\lambda_2$, $\lambda_3$, $\lambda_4$, $\lambda_5$, $\lambda_5^*$ \\ \hline
3 & CP2 & & & $\mu_1^2=\mu_2^2$, $\lambda_1=\lambda_2$, $\lambda_3$, $\lambda_4$, Re($\lambda_5$) \\ \hline
4 & U(1$)_{\text{}}$ & & & $\mu_1^2$, $\mu_2^2$, $\lambda_1$, $\lambda_2$, $\lambda_3$, $\lambda_4$\\ \hline
5 & CP1 $\otimes$ SO(2) & & & $\mu_1^2=\mu_2^2$, $\lambda_1=\lambda_2$, $\lambda_3$, $\lambda_4$, Re($\lambda_5)=2\lambda_1-\lambda_{34}$\\ \hline
6 & SU(2$)$& & & $\mu_1^2=\mu_2^2$, $\lambda_1=\lambda_2$, $\lambda_3$, $\lambda_4=2\lambda_1-\lambda_3$\\ \hline
7 &Sp(2$)_{\phi_1+\phi_2}$ & & & $\mu_1^2$, $\mu_2^2$, Re($m_{12}^2$), $\lambda_1$, $\lambda_2$, $\lambda_3$, $\lambda_4=\text{Re}(\lambda_5)$, Re($\lambda_6$), Re($\lambda_7$) \\ \hline
8 & (CP1 $\rtimes \,S_2)\otimes$ Sp(2$)_{\phi_1+\phi_2}$& & & $\mu_1^2=\mu_2^2$,  Re($m_{12}^2$), $\lambda_1=\lambda_2$, $\lambda_3$, $\lambda_4=\text{Re}(\lambda_5)$, Re($\lambda_6$)=Re($\lambda_7$) \\ \hline
9 & $(S_2\rtimes Z_2)\otimes$ Sp(2$)_{\phi_1+\phi_2}$ & & & $\mu_1^2=\mu_2^2$, $\lambda_1=\lambda_2$, $\lambda_3$, Re($\lambda_5$)$=\pm \lambda_4$\\ \hline
10 & U(1$)_{\text{}}$ $\otimes$ Sp(2$)_{\phi_1\phi_2}$ & & & $\mu_1^2=\mu_2^2$, $\lambda_1=\lambda_2={1\over 2}\lambda_3$, $\lambda_4$\\ \hline
11 & Sp(2)$_{\phi_1}\otimes$ Sp(2)$_{\phi_2}$ & & & $\mu_1^2$, $\mu_2^2$, $\lambda_1$, $\lambda_2$, $\lambda_3$\\ \hline
12 & $S_2\,\otimes$ Sp(2)$_{\phi_1}\otimes$ Sp(2)$_{\phi_2}$ & & & $\mu_1^2=\mu_2^2$, $\lambda_1=\lambda_2$, $\lambda_3$\\ \hline
13 & Sp(4) & & & $\mu_1^2=\mu_2^2$, $\lambda_1=\lambda_2={1\over 2}\lambda_3$\\ \hline
\end{tabular}
\caption{\it Parameter relations for the $13$ accidental symmetries related to the 2HDM potential~\cite{Pilaftsis:2011ed}. 
Note that all entries which contain symplectic group are only
custodially symmetric, and so violate U(1)$_Y$. The subscript ${\phi_1+\phi_2}$ shows an Sp(2) transformation that acts on both
$(\phi_1,
i \sigma^2 \phi_1^* 
)^{\mathsf{T}}$ and $
(
\phi_2,
i \sigma^2 \phi_2^* 
)^{\mathsf{T}}$. Additionally, the subscript ${\phi_1 \phi_2}$ denotes an Sp(2)
transformation acting on
$(
\phi_1,
i \sigma^2 \phi_2^* 
)^{\mathsf{T}}$. 
\label{tab1}}
\end{table}

Likewise, we may use maximal and minimal invariants to construct the
3HDM potentials. The symmetry groups and the 3HDM potentials as functions
of prime invariants are:
\begin{itemize}
\item {SO(2)$_{\phi_1,\phi_2}$}: $V[S_{11}+S_{22},D_{12}^2, T^2_{12},S_{33},
D_{13}^2+D_{23}^2, T^2_{13}+ T^2_{23}]$.
\item {SO(2)$_{\phi_1,\phi_2}\otimes$ Sp(2)$_{\phi_3}$}: $V[S_{11}+S_{22},D_{12}^2, T^2_{12},S_{33}]$.
\item SU(2)$_{\phi_1,\phi_2}$: $V[S_{11}+S_{22},D_{12}^2,S_{33},D_{13}^2+D_{23}^2]$.
\item SU(2)$_{\phi_1,\phi_2}\otimes$ Sp(2)$_{\phi_3}$: $V[S_{11}+S_{22},D_{12}^2,S_{33}]$.
\item Sp(2$)_{\phi_1+\phi_2+\phi_3}$: $V[S_{11},S_{22},S_{33},S_{12},S_{13},S_{23}]$.
\item Sp(2$)_{\phi_1+\phi_2}\otimes$ Sp(2)$_{\phi_3}$: $V[S_{11},S_{22},S_{33},S_{12}]$.
\item { Sp(2$)_{\phi_1 \phi_2}$ $\otimes$ Sp(2$)_{\phi_3}$}: $V[S_{11}+S_{22},D_{12}^{\prime 2},S_{33}]$.
\item { Sp(2$)_{\phi_1 \phi_2}$}: $V[S_{11}+S_{22},S_{33},D_{12}^{\prime 2},D_{13}^{\prime 2}+D_{23}^{\prime 2}]$.
\item Sp(2)$_{\phi_1}\otimes$Sp(2)$_{\phi_2}\otimes$Sp(2)$_{\phi_3}$: $V[S_{11},S_{22},S_{33}]$. 
\item SO(3): $V[S_{11}+S_{22}+S_{33}, D_{12}^2+D_{13}^2+D_{23}^2, T^2_{12}+T^2_{13}+ T^2_{23}]$.
\item Sp(4)$\otimes$ Sp(2)$_{\phi_3}$: $V[S_{11}+S_{22},S_{33}]$.
\item {SU(3)$\otimes$U(1)}: $V[S_{11}+S_{22}+S_{33},D_{12}^2+D_{13}^2+D_{23}^2]$.
\item Sp(6): $V[S_{11}+S_{22}+S_{33}]$.
\end{itemize}
In the above list, we have used the subscript ${\phi_1+\phi_2+\phi_3}$
to show a transformation that acts simultaneously on
$(
\phi_1,
i \sigma^2 \phi_1^* 
)^{\mathsf{T}}$,
$(
\phi_2,
i \sigma^2 \phi_2^* 
)^{\mathsf{T}}$
and 
$(
\phi_3,
i \sigma^2 \phi_3^* 
)^{\mathsf{T}}$. Also, the subscripts ${\phi_1 \phi_2}$ and ${\phi_3}$
denote Sp(2) transformation that act on 
$(
\phi_1,
i \sigma^2 \phi_2^* 
)^{\mathsf{T}}$
and
$(
\phi_3,
i \sigma^2 \phi_3^* 
)^{\mathsf{T}}$, respectively. Finally, the subscripts ${\phi_1, \phi_2}$ denote SU(2) or SO(2) transformations acting on
$(
\phi_1,
\phi_2
)^{\mathsf{T}}$.

\begin{table*}[t]
\small
\centering
\begin{tabular}{| c | c c | c l |}
\hline
No. & Symmetry & & & Non-zero parameters for 3HDM potentials\\
\hline \hline
\multirow{5}{*}{1} & \multirow{5}{*}{CP1} & & & 
$\mu^2_1$, $\mu^2_2$, $\mu^2_3$, Re($m_{12}^2)$, Re($m_{13}^2)$, Re($m_{23}^2)$, $\lambda_{11}$, $\lambda_{22}$, $\lambda_{33}$,\\
& & & & $\lambda_{1122}$, $\lambda_{1133} $, $\lambda_{2233}$, $\lambda_{1221}$, $\lambda_{1331}$, $\lambda_{2332}$, Re$(\lambda_{1212})$, Re$(\lambda_{1313})$, Re($\lambda_{2323})$, \\
& & & & $ \text{Re}( { \lambda_{1213}}) $, $\text{Re}({ \lambda_{2113}} )$, $\text{Re}( { \lambda_{1223}})$, $\text{Re}({\lambda_{2123}})$, $ \text{Re}({\lambda_{1323}} )$, $\text{Re} ({ \lambda_{1332}})$,\\
& & & &
$\text{Re}({\lambda_{1112}})$, $\text{Re}({ \lambda_{2212}})$, $ \text{Re}({ \lambda_{3312}})$, $\text{Re}({\lambda_{1113}})$, $\text{Re}({\lambda_{2213}})$, $\text{Re}( { \lambda_{3313}})$,\\
& & & &
$\text{Re}({\lambda_{1123}})$, $\text{Re}( { \lambda_{2223}})$, $ \text{Re}( { \lambda_{3323}})$
\\
\hline
\multirow{2}{*}{2} & \multirow{2}{*}{$Z_2$} & & & 
$\mu^2_1$, $\mu^2_2$, $\mu^2_3$, $\lambda_{11}$, $\lambda_{22}$, $\lambda_{33}$, $\lambda_{1122}$, $\lambda_{1133} $, $\lambda_{2233}$, $\lambda_{1221}$, $\lambda_{1331}$, $\lambda_{2332}$, \\
& & & & \{$m^2_{13}$, $\lambda_{1212}$, $\lambda_{1313}$, $\lambda_{2323}$, $\lambda_{1232}$, $\lambda_{1113}$, $\lambda_{2213}$, $\lambda_{3313}$ and ${\rm H.c.}$\}
\\
\hline
\multirow{2}{*}{2$^\prime$} & \multirow{2}{*}{$Z_2^{\prime}$} & & & 
$\mu^2_1$, $\mu^2_2$, $\mu^2_3$, $\lambda_{11}$, $\lambda_{22}$, $\lambda_{33}$, $\lambda_{1122}$, $\lambda_{1133} $, $\lambda_{2233}$, $\lambda_{1221}$, $\lambda_{1331}$, $\lambda_{2332}$,\\
& & & & \{$m^2_{23}$, $\lambda_{1212}$, $\lambda_{1313}$, $\lambda_{2323}$, $\lambda_{1213}$, $\lambda_{1123}$, $\lambda_{2223}$, $\lambda_{3323}$ and ${\rm H.c.}$\}
\\
\hline
\multirow{2}{*}{3} & \multirow{2}{*}{$Z_2\, \otimes \, Z_2^{\prime}$} & & & 
$\mu^2_1$, $\mu^2_2$, $\mu^2_3$, $\lambda_{11}$, $\lambda_{22}$, $\lambda_{33}$, $\lambda_{1122}$, $\lambda_{1133} $, $\lambda_{2233}$, $\lambda_{1221}$, $\lambda_{1331}$, $\lambda_{2332}$,\\
& & & & \{$\lambda_{1212}$, $\lambda_{1313}$, $\lambda_{2323}$ and ${\rm H.c.}$\}
\\
\hline
\multirow{2}{*}{4} & \multirow{2}{*}{$Z_3$} & & & $\mu^2_1$, $\mu^2_2$, $\mu^2_3$, $\lambda_{11}$, $\lambda_{22}$, $\lambda_{33}$, $\lambda_{1122}$, $\lambda_{1133}$, $\lambda_{2233}$, $\lambda_{1221}$, $\lambda_{1331}$, $\lambda_{2332}$,\\
& & & & \{$\lambda_{1213}$, $\lambda_{1323}$, $\lambda_{2123}$ and ${\rm H.c.}$\}\\
\hline
\multirow{2}{*}{5} & \multirow{2}{*}{$Z_4$} & & & $\mu^2_1$, $\mu^2_2$, $\mu^2_3$, $\lambda_{11}$, $\lambda_{22}$, $\lambda_{33}$, $\lambda_{1122}$, $\lambda_{1133}$, $\lambda_{2233}$, $\lambda_{1221}$, $\lambda_{1331}$, $\lambda_{2332}$,\\
& & & & \{$\lambda_{1212}$, $\lambda_{1323}$ and ${\rm H.c.}$\}\\
\hline
\multirow{2}{*}{5$^\prime$} & \multirow{2}{*}{$Z_4^\prime$} & & & $\mu^2_1$, $\mu^2_2$, $\mu^2_3$, $\lambda_{11}$, $\lambda_{22}$, $\lambda_{33}$, $\lambda_{1122}$, $\lambda_{1133}$, $\lambda_{2233}$, $\lambda_{1221}$, $\lambda_{1331}$, $\lambda_{2332}$,\\
& & & & \{$\lambda_{1313}$, $\lambda_{3212}$ and ${\rm H.c.}$\}\\
\hline
\multirow{2}{*}{6} & \multirow{2}{*}{$^a$U(1$)_{}$} & & & $\mu^2_1$, $\mu^2_2$, $\mu^2_3$, $\lambda_{11}$, $\lambda_{22}$, $\lambda_{33}$, $\lambda_{1122}$, $\lambda_{1133}$, $\lambda_{2233}$, $\lambda_{1221}$, $\lambda_{1331}$, $\lambda_{2332}$,\\
& & & & \{$\lambda_{1323}$ and ${\rm H.c.}$\}\\
\hline
\multirow{2}{*}{6$^\prime$} & \multirow{2}{*}{$^b$U(1$)_{}^{\prime} $} & & & $\mu^2_1$, $\mu^2_2$, $\mu^2_3$, $\lambda_{11}$, $\lambda_{22}$, $\lambda_{33}$, $\lambda_{1122}$, $\lambda_{1133}$, $\lambda_{2233}$, $\lambda_{1221}$, $\lambda_{1331}$, $\lambda_{2332}$,\\
& & & & \{$m^2_{12}$, $\lambda_{1212}$, $\lambda_{1112}$, $\lambda_{2212}$, $\lambda_{3312}$, $\lambda_{1332}$ and ${\rm H.c.}$\}\\
\hline
\multirow{1}{*}{7} & \multirow{1}{*}{U(1$)_{\text{}}$ $\otimes$ U(1$)_{\text{}}^{\prime} $} & & & $\mu^2_1$, $\mu^2_2$, $\mu^2_3$, $\lambda_{11}$, $\lambda_{22}$, $\lambda_{33}$, $\lambda_{1122}$, $\lambda_{1133}$, $\lambda_{2233}$, $\lambda_{1221}$, $\lambda_{1331}$, $\lambda_{2332}$\\
\hline
\multirow{2}{*}{8} & \multirow{2}{*}{$Z_2\, \otimes \, ${U(1$)_{}^{\prime} $}} & & & $\mu^2_1$, $\mu^2_2$, $\mu^2_3$, $\lambda_{11}$, $\lambda_{22}$, $\lambda_{33}$, $\lambda_{1122}$, $\lambda_{1133}$, $\lambda_{2233}$, $\lambda_{1221}$, $\lambda_{1331}$, $\lambda_{2332}$,\\
& & & &\{{$\lambda_{1212}$ and ${\rm H.c.}$}\}\\
\hline
\multirow{2}{*}{9} & \multirow{2}{*}{CP1\,$\otimes$ Sp(2)$_{\phi_3}$} & & & 
$\mu^2_1$, $\mu^2_2$, $\mu^2_3$, Re($m_{12}^2)$, $\lambda_{11}$, $\lambda_{22}$, $\lambda_{33}$, $\lambda_{1122}$, $\lambda_{1133} $, $\lambda_{2233}$, $\lambda_{1221}$, Re$(\lambda_{1212})$, \\
& & & & 
$\text{Re}({\lambda_{1112}})$, $\text{Re}({ \lambda_{2212}})$, $\text{Re}({ \lambda_{3312}})$
\\
\hline
\multirow{1}{*}{10} & \multirow{1}{*}{CP1\,$\otimes \, Z_2 \, \otimes$ Sp(2)$_{\phi_3}$} & & & 
$\mu^2_1$, $\mu^2_2$, $\mu^2_3$, $\lambda_{11}$, $\lambda_{22}$, $\lambda_{33}$, $\lambda_{1122}$, $\lambda_{1133} $, $\lambda_{2233}$, $\lambda_{1221}$, Re$(\lambda_{1212})$
\\
\hline
\multirow{1}{*}{11} & \multirow{1}{*}{U(1$)$ $\otimes$ Sp(2)$_{\phi_3}$} & & & $\mu^2_1$, $\mu^2_2$, $\mu^2_3$, $\lambda_{11}$, $\lambda_{22}$, $\lambda_{33}$, $\lambda_{1122}$, $\lambda_{1133}$, $\lambda_{2233}$, $\lambda_{1221}$\\
\hline
\multirow{2}{*}{12} & \multirow{2}{*}{CP2} & & & $\mu^2_1=\mu^2_2$, $\mu^2_3$, $\lambda_{11}=\lambda_{22}$, $\lambda_{33}$, $\lambda_{1122}$, $\lambda_{1221}$, $\lambda_{1133}=\lambda_{2233}$, $\lambda_{1331}=\lambda_{2332}$, \\
& & & & 
 Re($\lambda_{1313}$)=Re($\lambda_{2323}$), Re($\lambda_{1212}$), 
\{$\lambda_{1112}=-\lambda_{2212}$ and ${\rm H.c.}$\}
\\
\hline
\multirow{2}{*}{13} & \multirow{2}{*}{CP2 $\otimes$ Sp(2)$_{\phi_3}$} & & & $\mu^2_1=\mu^2_2$, $\mu^2_3$, $\lambda_{11}=\lambda_{22}$, $\lambda_{33}$, $\lambda_{1122}$, $\lambda_{1221}$, $\lambda_{1133}=\lambda_{2233}$, Re($\lambda_{1212}$), \\
& & & & 
\{$\lambda_{1112}=-\lambda_{2212}$ and ${\rm H.c.}$\}
\\
\hline
\multirow{2}{*}{14} & \multirow{2}{*}{SO(2)$_{\phi_1,\phi_2}$} & & & $\mu^2_1=\mu^2_2$, $\mu^2_3$, $\lambda_{11}=\lambda_{22}$, $\lambda_{33}$, $\lambda_{1122}$, $\lambda_{1221}$, $\lambda_{1133}=\lambda_{2233}$, $\lambda_{1331}=\lambda_{2332}$, \\
& & & & Re($\lambda_{1313})=\text{Re}(\lambda_{2323}$), Re($\lambda_{1212})=2\lambda_{11}-(\lambda_{1122}+\lambda_{1221})$, \\
\hline
\multirow{2}{*}{15} & \multirow{2}{*}{$D_3$} & & & $\mu^2_1=\mu^2_2$, $\mu^2_3$, $\lambda_{11}=\lambda_{22}$, $\lambda_{33}$, $\lambda_{1122}$, $\lambda_{1133}=\lambda_{2233}$, $\lambda_{1221}$, $\lambda_{1331}=\lambda_{2332}$,\\
& & & &\{$\lambda_{2131}=-\lambda_{1232}$, $\lambda_{1323}$ and ${\rm H.c.}$\}\\
\hline
\multirow{2}{*}{16} & \multirow{2}{*}{$D_4$} & & & $\mu^2_1=\mu^2_2$, $\mu^2_3$, $\lambda_{11}=\lambda_{22}$, $\lambda_{33}$, $\lambda_{1122}$, $\lambda_{1133}=\lambda_{2233}$, $\lambda_{1221}$, \{$\lambda_{1212}$ and ${\rm H.c.}$\}, \\
& & & &$\lambda_{1331}=\lambda_{2332}=\text{Re}(\lambda_{3231})$\\
\hline
\multirow{1}{*}{17} & \multirow{1}{*}{$D_3\,\otimes$ Sp(2)$_{\phi_3}$} & & & 
$\mu^2_1=\mu^2_2$, $\mu^2_3$, $\lambda_{11}=\lambda_{22}$, $\lambda_{33}$, $\lambda_{1122}$, $\lambda_{1133}=\lambda_{2233}$, $\lambda_{1221}$
\\
\hline
\multirow{1}{*}{18} & \multirow{1}{*}{$D_4\,\otimes$ Sp(2)$_{\phi_3}$} & & & $\mu^2_1=\mu^2_2$, $\mu^2_3$, $\lambda_{11}=\lambda_{22}$, $\lambda_{33}$, $\lambda_{1122}$, $\lambda_{1133}=\lambda_{2233}$, $\lambda_{1221}$, Re$(\lambda_{1212})$\\
\hline
\end{tabular}
\end{table*}
\begin{table*}[!h]
\small
\begin{tabular}{| c | c c | c l |}
\hline
No. & Symmetry & & & Non-zero parameter space of 3HDM\\
\hline \hline
\multirow{2}{*}{19} & \multirow{2}{*}{                                                                                                                                                                                                                                                                                                                                                                                                                                                                                                                                                                                                                                                                                                                                                                                                                                                                                                                                                                                                                                                                                                                                                                                                                                                                                                                                                                                                                                                                                                                                                                                                                                                                                                                                                                                                                                                                                                                                                                                                                                                                                                             SO(2)$_{\phi_1,\phi_2}\otimes$ Sp(2)$_{\phi_3}$} & & & $\mu^2_1=\mu^2_2$, $\mu^2_3$, $\lambda_{11}=\lambda_{22}$, $\lambda_{33}$, $\lambda_{1122}$, $\lambda_{1133}=\lambda_{2233}$, \\
& & & & $\lambda_{1221}= \text{Re}(\lambda_{1212})=\lambda_{11}-{1\over2}\lambda_{1122}$\\
\hline
\multirow{2}{*}{20} & \multirow{2}{*}{SU(2)$_{\phi_1,\phi_2}$} & & & $\mu^2_1=\mu^2_2$, $\mu^2_3$, $\lambda_{11}=\lambda_{22}$, $\lambda_{33}$, $\lambda_{1122}=2\lambda_{11}-\lambda_{1221}$, $\lambda_{1221}$,\\
& & & &$\lambda_{1133}=\lambda_{2233}$, $\lambda_{1331}=\lambda_{2332}$\\
\hline
\multirow{1}{*}{21} & \multirow{1}{*}{SU(2)$_{\phi_1,\phi_2}\otimes$ Sp(2)$_{\phi_3}$} & & & $\mu^2_1=\mu^2_2$, $\mu^2_3$, $\lambda_{11}=\lambda_{22}$, $\lambda_{33}$, $\lambda_{1122}=2\lambda_{11}-\lambda_{1221}$, $\lambda_{1133}=\lambda_{2233}$,\\
& & & & $\lambda_{1221}$\\
\hline
\multirow{5}{*}{22} & \multirow{5}{*}{Sp(2$)_{\phi_1+\phi_2+\phi_3}$} & & & 
$\mu^2_1$, $\mu^2_2$, $\mu^2_3$, Re($m_{12}^2)$, Re($m_{13}^2)$, Re($m_{23}^2)$, $\lambda_{11}$, $\lambda_{22}$, $\lambda_{33}$, \\
& & & & $\lambda_{1122}$, $\lambda_{1133} $, $\lambda_{2233}$, $\lambda_{1221}$= Re$(\lambda_{1212})$, $\lambda_{1331}$=Re$(\lambda_{1313})$, \\
& & & &
$\lambda_{2332}$=Re($\lambda_{2323})$, $\text{Re}( { \lambda_{1213}}) = \text{Re}({ \lambda_{2113}} )$, $\text{Re}( { \lambda_{1223}}) = \text{Re}({\lambda_{2123}})$,\\
& & & &
$ \text{Re}( { \lambda_{1323}} ) = \text{Re}({ \lambda_{1332}})$, $\text{Re}({ \lambda_{1112}}), \text{Re}({\lambda_{2212}}), \text{Re}({\lambda_{3312}})$,\\
& & & &
$\text{Re}({\lambda_{1113}}), \text{Re}({ \lambda_{2213}}), \text{Re}({\lambda_{3313}})$, $\text{Re}({\lambda_{1123}}), \text{Re}({ \lambda_{2223}}), \text{Re}({\lambda_{3323}})$
\\
\hline
\multirow{2}{*}{23} & \multirow{2}{*}{$Z_2 \, \otimes$ Sp(2$)_{\phi_1+\phi_2+\phi_3}$} & & & 
$\mu^2_1$, $\mu^2_2$, $\mu^2_3$, Re($m_{13}^2)$, $\lambda_{11}$, $\lambda_{22}$, $\lambda_{33}$, $\lambda_{1122}$, $\lambda_{1133} $, $\lambda_{2233}$, \\
& & & & $\lambda_{1221}$= Re$(\lambda_{1212})$, $\lambda_{1331}$=Re$(\lambda_{1313})$, $\lambda_{2332}$=Re($\lambda_{2323})$, \\
& & & &
$\text{Re}({\lambda_{1113}}), \text{Re}({ \lambda_{2213}}), \text{Re}({\lambda_{3313}})$
\\
\hline
\multirow{2}{*}{23$^\prime$} & \multirow{2}{*}{$Z_2^{\prime}\,\otimes \,$Sp(2$)_{\phi_1+\phi_2+\phi_3}$} & & & 
$\mu^2_1$, $\mu^2_2$, $\mu^2_3$, Re($m_{23}^2)$, $\lambda_{11}$, $\lambda_{22}$, $\lambda_{33}$, $\lambda_{1122}$, $\lambda_{1133} $, $\lambda_{2233}$, \\
& & & & $\lambda_{1221}$= Re$(\lambda_{1212})$, $\lambda_{1331}$=Re$(\lambda_{1313})$, $\lambda_{2332}$=Re($\lambda_{2323})$, \\
& & & &
$\text{Re}({\lambda_{1123}}), \text{Re}({ \lambda_{2223}}), \text{Re}({\lambda_{3323}})$
\\
\hline
\multirow{2}{*}{24} & \multirow{2}{*}{$Z_2 \,\otimes \, Z_2^{\prime}\otimes$ Sp(2$)_{\phi_1+\phi_2+\phi_3}$} & & & 
$\mu^2_1$, $\mu^2_2$, $\mu^2_3$, $\lambda_{11}$, $\lambda_{22}$, $\lambda_{33}$, $\lambda_{1122}$, $\lambda_{1133} $, $\lambda_{2233}$, \\
& & & & $\lambda_{1221}$= Re$(\lambda_{1212})$, $\lambda_{1331}$=Re$(\lambda_{1313})$, $\lambda_{2332}$=Re($\lambda_{2323})$\\
\hline
\multirow{1}{*}{25} & \multirow{1}{*}{$Z_4 \, \otimes \,$ Sp(2$)_{\phi_1+\phi_2+\phi_3}$} & & & $\mu^2_1$, $\mu^2_2$, $\mu^2_3$, $\lambda_{11}$, $\lambda_{22}$, $\lambda_{33}$, $\lambda_{1122}$, $\lambda_{1133} $, $\lambda_{2233}$, $\lambda_{1221}$= Re$(\lambda_{1212})$
\\
\hline
\multirow{6}{*}{26} & \multirow{6}{*}{(CP1 $\rtimes S_2)\otimes$ Sp(2$)_{\phi_1+\phi_2+\phi_3}$}& & & 
$\mu^2_1=\mu^2_2$, $\mu^2_3$, Re($m_{12}^2)$, Re($m_{13}^2)$=Re($m_{23}^2)$, $\lambda_{11}=\lambda_{22}$, $\lambda_{33}$, \\
& & & &
$\lambda_{1122}$, $\lambda_{1133} =\lambda_{2233}$, $\lambda_{1221}$= Re$(\lambda_{1212})$, \\
& & & & 
$\lambda_{1331}$=$\lambda_{2332}$=Re$(\lambda_{1313})$=Re($\lambda_{2323})$, $\text{Re}( { \lambda_{1323}} ) = \text{Re} ({ \lambda_{1332}})$, \\
& & & &
$\text{Re}( { \lambda_{1223}}) = \text{Re}({\lambda_{2123}})$=$ \text{Re}( { \lambda_{1213}}) = \text{Re}({ \lambda_{2113}} )$,\\
& & & &
$\text{Re}( { \lambda_{3313}})= \text{Re}({\lambda_{3323}})$, $\text{Re}({\lambda_{1112}})=\text{Re}({ \lambda_{2212}})$,\\
& & & &
 $\text{Re}({ \lambda_{3312}})$, 
 $\text{Re}({\lambda_{1113}})=\text{Re}({\lambda_{1123}})=\text{Re}({\lambda_{2213}})=\text{Re}({\lambda_{2223}})$
\\
\hline
\multirow{2}{*}{27} & \multirow{2}{*}{$D_4\,\otimes$ Sp(2$)_{\phi_1+\phi_2+\phi_3}$}& & & 
$\mu^2_1=\mu^2_2$, $\mu^2_3$, $\lambda_{11}=\lambda_{22}={1 \over 2}\lambda_{1122}$, $\lambda_{33}$, $\lambda_{1133}=\lambda_{2233}$, \\
& & & &$\lambda_{1221}=$Re($\lambda_{1212}$)
\\
\hline
\multirow{2}{*}{28} & \multirow{2}{*}{Sp(2$)_{\phi_1+\phi_2}$ $\otimes$ Sp(2$)_{\phi_3}$}& & & 
$\mu^2_1$, $\mu^2_2$, $\mu^2_3$, Re($m_{12}^2)$, $\lambda_{11}$, $\lambda_{22}$, $\lambda_{33}$, $\lambda_{1122}$, $\lambda_{1133} $, $\lambda_{2233}$, \\
& & & &
 $\lambda_{1221}$= Re$(\lambda_{1212})$, $\text{Re}({ \lambda_{1112}}), \text{Re}({\lambda_{2212}}), \text{Re}({\lambda_{3312}})$\\
\hline
\multirow{2}{*}{29} & \multirow{2}{*}{Sp(2$)_{\phi_1 \phi_2}$} & & & 
$\mu^2_1=\mu^2_2$, $\mu^2_3$, $\lambda_{11}=\lambda_{22}={1 \over 2}\lambda_{1122} $, $\lambda_{33}$, $\lambda_{1133}=\lambda_{2233}$, 
$\lambda_{1221},$ \\
& & & & $\lambda_{1331}=\lambda_{2332}$\\
\hline
\multirow{1}{*}{30} & \multirow{1}{*}{Sp(2$)_{\phi_1 \phi_2}$ $\otimes$ Sp(2$)_{\phi_3}$} & & & 
$\mu^2_1=\mu^2_2$, $\mu^2_3$, $\lambda_{11}=\lambda_{22}={1 \over 2}\lambda_{1122} $, $\lambda_{33}$, $\lambda_{1133}=\lambda_{2233}$, 
$\lambda_{1221}$ \\
\hline
\multirow{2}{*}{31} & \multirow{2}{*}{$A_4$} & & & $\mu^2_1=\mu^2_2=\mu^2_3$, $\lambda_{11}=\lambda_{22}=\lambda_{33}$, 
$\lambda_{1122}=\lambda_{1133}=\lambda_{2233}$,\\
& & & & $\lambda_{1221}=\lambda_{1331}=\lambda_{2332}$, \{$\lambda_{1212}=\lambda_{1313}=\lambda_{2323}$, and ${\rm H.c.}$\}\\
\hline
\multirow{2}{*}{32} & \multirow{2}{*}{$S_4$} & & & $\mu^2_1=\mu^{2}_2=\mu^{2}_3$, $\lambda_{11}=\lambda_{22}=\lambda_{33}$, $\lambda_{1122}=\lambda_{1133}=\lambda_{2233}$, \\
& & & &$\lambda_{1221}=\lambda_{1331}=\lambda_{2332}$, $\text{Re}(\lambda_{1212})=\text{Re}(\lambda_{1313})=\text{Re}(\lambda_{2323})$\\
\hline
\multirow{3}{*}{33} & \multirow{3}{*}{SO(3)} & & & $\mu^2_1=\mu^{2}_2=\mu^{2}_3$, $\lambda_{11}=\lambda_{22}=\lambda_{33}$, $\lambda_{1122}=\lambda_{1133}=\lambda_{2233}$, \\
& & & &$\lambda_{1221}=\lambda_{1331}=\lambda_{2332}$, \\
& & & & \text{Re}($\lambda_{1212})=\text{Re}(\lambda_{1313})=\text{Re}(\lambda_{2323})=2\lambda_{11}-(\lambda_{1122}+\lambda_{1221})$\\
\hline
\end{tabular}
\end{table*}

\begin{table}[!h]
\small
\begin{tabular}{| c | c c | c l |}
\hline
No. & Symmetry & & & Non-zero parameter space of 3HDM\\
\hline \hline
\multirow{2}{*}{34} & \multirow{2}{*}{$S_4\, \otimes$ Sp(2$)_{\phi_1+\phi_2+\phi_3}$} & & & $\mu^2_1=\mu^{2}_2=\mu^{2}_3$, $\lambda_{11}=\lambda_{22}=\lambda_{33}={1 \over 2} \lambda_{1122}={1 \over 2} \lambda_{1133}={1 \over 2}\lambda_{2233}$, \\
& & & &$\lambda_{1221}=\lambda_{1331}=\lambda_{2332}$=$\text{Re}(\lambda_{1212})=\text{Re}(\lambda_{1313})=\text{Re}(\lambda_{2323})$\\
\hline
\multirow{3}{*}{35} & \multirow{3}{*}{$\Delta(54)$} & & & $\mu^2_1=\mu^2_2=\mu^2_3$, $\lambda_{11}=\lambda_{22}=\lambda_{33}$, 
$\lambda_{1122}=\lambda_{1133}=\lambda_{2233}$,\\
& & & & $\lambda_{1221}=\lambda_{1331}=\lambda_{2332}$, \{$\lambda_{1213}=\lambda_{2123}=\lambda_{3231}$ and ${\rm H.c.}$\}\\
\hline
\multirow{3}{*}{36} & \multirow{3}{*}{$\Sigma(36)$} & & & $\mu^2_1=\mu^{2}_2=\mu^{2}_3$, $\lambda_{11}=\lambda_{22}=\lambda_{33}$, $\lambda_{1122}=\lambda_{1133}=\lambda_{2233}$, \\
& & & &$\lambda_{1221}=\lambda_{1331}=\lambda_{2332}$, $\text{Re}(\lambda_{1213})=\text{Re}(\lambda_{1323})=\text{Re}(\lambda_{1232})$= \\
& & & &${3\over 4}(2\lambda_{11}-\lambda_{1122}-\lambda_{1221})$\\
\hline
\multirow{1}{*}{37} & Sp(2)$_{\phi_1}\,\otimes$ Sp(2)$_{\phi_2}\,\otimes$ Sp(2)$_{\phi_3}$& & & $\mu^2_1$, $\mu^2_2$, $\mu^2_3$, $\lambda_{11}$, $\lambda_{22}$, $\lambda_{33}$, $\lambda_{1122}$, $\lambda_{1133}$, $\lambda_{2233}$\\
\hline
\multirow{1}{*}{38} & Sp(4) $\otimes$ Sp(2)$_{\phi_3}$ & & & $\mu^2_1=\mu^2_2$, $\mu^2_3$, $\lambda_{11}=\lambda_{22}={1 \over 2}\lambda_{1122} $, $\lambda_{33}$, $\lambda_{1133}=\lambda_{2233}$
\\
\hline
\multirow{2}{*}{39} & \multirow{2}{*}{SU(3) $\otimes$ U(1)} & & & $\mu^{2}_1=\mu^{2}_2=\mu^{2}_3$, $\lambda_{11}=\lambda_{22}=\lambda_{33}$, $\lambda_{1122}=\lambda_{1133}=\lambda_{2233}$,\\
& & & & $\lambda_{1221}=\lambda_{1331}=\lambda_{2332}=2 \lambda_{11}-\lambda_{1122}$\\
\hline
\multirow{1}{*}{40} & Sp(6) & & & $\mu^{2}_1=\mu^{2}_2=\mu^{2}_3$, $\lambda_{11}=\lambda_{22}=\lambda_{33}={1 \over 2}\lambda_{1122}={1 \over 2}\lambda_{1133}={1 \over 2}\lambda_{2233}$\qquad\qquad\\
\hline
\end{tabular}
\caption{\it Non-zero parameters for the $40$ accidental symmetries related to the 3HDM~potential. Note that all entries which contain symplectic group are only custodially symmetric, and  so violate U(1)$_Y$.
$^{a,b}$The generators of \text{U}(1) and U(1)$^\prime$ are ${\rm diag}(e^{i\alpha},e^{-i\alpha},1)$ and ${\rm diag}(e^{i\beta/3},e^{i\beta/3},e^{-i2\beta/3})$, with $\alpha, \beta \in [0,2\pi)$, respectively. 
Moreover, the subscripts ${\phi_1, \phi_2}$ denote SU(2) or SO(2) transformations acting on
$(
\phi_1,
\phi_2
)^{\mathsf{T}}$ and the subscript ${\phi_1+\phi_2+\phi_3}$ shows an Sp(2) transformation acting on all doublets
$(
\phi_1,
i \sigma^2 \phi_1^* 
)^{\mathsf{T}},$
$(
\phi_2,
i \sigma^2 \phi_2^* 
)^{\mathsf{T}}$ and 
$(
\phi_3,
i \sigma^2 \phi_3^* 
)^{\mathsf{T}}$. Finally, the subscripts ${\phi_1 \phi_2}$ and ${\phi_3}$ indicate an Sp(2) transformation acting on
$(
\phi_1,
i \sigma^2 \phi_2^* 
)^{\mathsf{T}}$
and
$(
\phi_3,
i \sigma^2 \phi_3^* 
)^{\mathsf{T}}$, respectively. 
}
\label{tab2}
\end{table}

\section{Irreducible representations of non-Abelian discrete symmetries}
\label{D}

The most familiar non-Abelian subgroups of SU(3) are summarized in
Section \ref{sec:ds}.  The direct sum decomposition of 3$\otimes 3$
tensor products in terms of irreducible representations of these
groups are listed below:
\begin{enumerate}
\item $1\oplus 1\oplus1\oplus1\oplus1\oplus1\oplus1\oplus1\oplus1: \Delta(3N^2)$
\item $1\oplus 1\oplus1\oplus1\oplus1\oplus1\oplus2: \Sigma(2N^2)$ 
\item $1 \oplus 1 \oplus 1 \oplus 3 \oplus 3: A_4$
\item $1\oplus2\oplus2\oplus2\oplus2: \Delta(6N^2)$
\item $1\oplus2\oplus3\oplus3: S_4$
\item $1\oplus2\oplus6: \Delta(6N^2)$
\item $1\oplus3\oplus5: A_5 $
\item $1\oplus4\oplus4: \Sigma (36\phi) $
\item $1\oplus8: \Sigma (72\phi), \Sigma (168), \Sigma (216\phi), \Sigma (360\phi) $
\item $3\oplus3\oplus3: \Sigma (36\phi), \Delta(3N^2), \Delta(6N^2) $
\item $3\oplus6: \Sigma(72\phi), \Sigma (168), \Sigma (216\phi), \Sigma (360\phi), \Delta(6N^2)$
\item $4\oplus5: A_5 $
\item $9: \Sigma (216\phi), \Sigma (360\phi)$
\end{enumerate}
As can be seen from the above list, only the decompositions (1)--(9)
are relevant for building SU(2)$_L$-preserving $n$HDM invariant potentials,
since they contain a singlet. However, only those
symmetries for which their prime factor decomposition lead to distinct
$n$HDM potentials can be considered as candidates for a novel symmetry of a
model. For example, the $A_5$ symmetry with $Z_n$ prime factors 2, 3,
5 does not lead to a distinct 3HDM potential. Moreover, in the case of
the 3HDM, the decompositions given in (6) and (9) produce potentials
invariant under SU(3). The remaining possibilities (10)--(13) have
been checked up to $Z_n$ prime factors of the model, but they do not
seem to lead to new forms of symmetric potentials.

For the case of the 3HDM, there are two non-Abelian discrete symmetries as
subgroups of SO(3), namely $D_3$ and $D_4$.  In addition, there are
non-Abelian discrete symmetries as subgroups of
SU(3)~\cite{Ishimori:2012zz,Ivanov:2012fp,Keus:2013hya,Ivanov:2014doa,deMedeirosVarzielas:2019rrp},
\begin{equation}
A_4, \, S_4, \, \left\{ \Sigma(18), \, \Delta(27), \, \Delta(54) \right\}, \, \Sigma(36).
\end{equation}

In Section \ref{sec:ds}, we have shown the procedure for constructing 3HDM potentials invariant under the non-Abelian discrete symmetries $D_3$ and $A_4$. Here, we apply this method for the rest of non-Abelian discrete symmetries of the 3HDM potential \cite{Ishimori:2012zz,Ivanov:2012fp,Keus:2013hya,Ivanov:2014doa,deMedeirosVarzielas:2019rrp}.

The other two-dimensional non-Abelian discrete symmetry of the 3HDM
is $D_4$. The $2\otimes 2$ tensor product of $D_4$ decomposes as
\begin{eqnarray}
D_4: \quad \textbf{2} \otimes \textbf{2}=\textbf{1} \oplus \textbf{1$^{\prime}$} \oplus \textbf{1$^{\prime\prime}$} \oplus \textbf{1$^{\prime\prime\prime}$}.
\end{eqnarray}
This group is generated by two generators,
\begin{equation}
\Delta_{D_4}^1=\text{diag}[{g_1} \otimes \sigma^0,{g_1}^* \otimes \sigma^0], \quad
\Delta_{D_4}^2=\text{diag}[{g_2} \otimes \sigma^0,{g_2}^* \otimes \sigma^0], 
\end{equation}
with
\begin{eqnarray} \small
g_1 = 
	\begin{pmatrix}
	i & 0 & 0 \\
	0 & -i & 0 \\
	0 & 0 & 1
	\end{pmatrix},
\quad
g_2 = 
	\begin{pmatrix}
	0 & -1 & 0 \\
	-1 & 0 & 0 \\
	0 & 0 & 1
	\end{pmatrix},
\end{eqnarray}
that satisfy the conditions: $g_1^4=1$, 
$g_2^2=1$ and $g_2 \cdot g_1 \cdot g_2= g_1^{-1}$.
The irreducible representations of $D_4$ in the bilinear-space may be given by \vspace{-0.2in}
\begin{flalign}
\begin{minipage}{0.45\textwidth}
\begin{eqnarray} \small
\textbf{1} &:& \;
	\begin{pmatrix}
	\phi_1^{\dagger} \phi_1 +\phi_2^{\dagger} \phi_2+\phi_3^{\dagger} \phi_3 
	\end{pmatrix},
\nonumber \\
\textbf{1$^{\prime}$} &:& \;
	\begin{pmatrix}
	\phi_1^{\dagger} \phi_1 - \phi_2^{\dagger} \phi_2
	\end{pmatrix},
\nonumber \\
	\textbf{1$^{\prime\prime}$} &:& \;
	\begin{pmatrix}
	{1\over\sqrt{3}}[\phi_1^{\dagger} \phi_1 +\phi_2^{\dagger} \phi_2-2 \phi_3^{\dagger} \phi_3]
	\end{pmatrix},
\nonumber \\
\textbf{1$^{\prime\prime\prime}$} &:& \;
	\begin{pmatrix}
	\phi_1^{\dagger} \phi_2 +\phi_2^{\dagger} \phi_1
	\end{pmatrix},
\nonumber \\
\textbf{1$^{\prime\prime\prime\prime}$} &:& \;
	\begin{pmatrix}
	-i [\phi_1^{\dagger} \phi_2 -\phi_2^{\dagger} \phi_1]
	\end{pmatrix},
\nonumber
\end{eqnarray}
\end{minipage}
\begin{minipage}{0.4\textwidth}
\begin{eqnarray}\small
\hspace{-0.1in}
\textbf{2} &:& \;
	\begin{pmatrix}
	\phi_1^{\dagger} \phi_3 + \phi_3^{\dagger} \phi_1 
	+\phi_2^{\dagger} \phi_3+\phi_3^{\dagger} \phi_2 \\
	-i [\phi_1^{\dagger} \phi_3 -\phi_3^{\dagger} \phi_1] 
	+i [\phi_2^{\dagger} \phi_3 -\phi_3^{\dagger} \phi_2] 
	\end{pmatrix}.
\nonumber
\end{eqnarray}
\end{minipage}
\nonumber \\
\end{flalign}
Hence, the $D_4$-invariant 3HDM potential takes on the form:
\begin{eqnarray} \small
V_{D_4}=&& 
-M_1\,{\bf{ 1}} -M_2\,{\bf {1^{\prime\prime}}}+
	 \Lambda_{0}\,{\bf{1^{\mathsf{2}}}}
    +\Lambda_{1}\,{\bf{{1^{\prime}}^\mathsf{2}}}
    +\Lambda_{2}\,{\bf{{1^{\prime\prime}}^\mathsf{2}}}
    +\Lambda_{3}\,{\bf{{1^{\prime\prime\prime}}^\mathsf{2}}}
 \nonumber \\    
 &&
   +\Lambda_{4}\,{\bf{{1^{\prime\prime\prime\prime}}^\mathsf{2}}}
       +\Lambda_{5}\,{\bf{{1}}} \cdot {\bf{{1^{\prime\prime}}}}
    +\Lambda_{6}\,{\bf{2^\mathsf{T}}\cdot \bf{2}}.
\end{eqnarray}
This potential can also be written as
\begin{align}
V_{D_4} = &
- \mu_1^2 \left( |\phi_1|^2
+|\phi_2|^2 \right)
- \mu_3^2 |\phi_3|^2
+\lambda_{11} \left( |\phi_1|^4 
+ |\phi_2|^4 \right)
+\lambda_{33} |\phi_3|^4
 \nonumber \\ &
+ \lambda_{1122} |\phi_1|^2 |\phi_2|^2
+ \lambda_{1133} \left(|\phi_1|^2 |\phi_3|^2
 +|\phi_2|^2 |\phi_3|^2 \right)
 \nonumber \\ &
+ \lambda_{1221} |\phi_1^{\dagger} \phi_2|^2
+ \lambda_{1212} (\phi_1^{\dagger} \phi_2)^2
 \nonumber \\ &
+ \lambda_{1331} \left( 
  |\phi_1^{\dagger} \phi_3|^2
+ |\phi_2^{\dagger} \phi_3|^2
+ |\phi_3^{\dagger} \phi_2|^2
+ (\phi_2^{\dagger} \phi_3) (\phi_1^{\dagger} \phi_3)
+(\phi_3^{\dagger} \phi_2) (\phi_3^{\dagger} \phi_1) \right),
\end{align}
where all parameters are real. 

Likewise, we find the $S_4$ invariant 3HDM potential. The $S_4$ group can be defined by the two generators~\cite{Pakvasa:1978tx,Yamanaka:1981pa,Brown:1984mq,Ishimori:2012zz},
\begin{equation}
\Delta_{S_4}^1=\sigma^0 \otimes {g_1} \otimes \sigma^0, \quad
\Delta_{S_4}^2=\sigma^0 \otimes {g_2} \otimes \sigma^0,
\end{equation}
with
\begin{eqnarray} \small
g_1 = 
	\begin{pmatrix}
	0 & 1 & 0 \\
	0 & 0 & 1\\
	1 & 0 & 0
	\end{pmatrix},
\quad
g_2 =
	\begin{pmatrix}
	0 & -1 & 0 \\
	-1 & 0 & 0 \\
	0 & 0 & 1
	\end{pmatrix},
\end{eqnarray}
which obey the conditions: $g_1^4=g_2^3=1$ and $g_1 \cdot g_2^2 \cdot g_1 = g_2$. 
The $\textbf{3} \otimes \textbf{3}$ tensor product of $S_4$ consists of one singlet \textbf{1}, one doublet \textbf{2} and two triplets, \textbf{3} and \textbf{3$^{\prime}$}
\begin{eqnarray}
S_4: \quad \textbf{3} \otimes \textbf{3}=\textbf{1} \oplus \textbf{2} \oplus \textbf{3} \oplus \textbf{3$^{\prime}$}.
\end{eqnarray}
The $S_4$-symmetric blocks {\bf{1}}, {\bf{2}}, {\bf{3}} and {\bf{3$^{\prime}$}} may be obtained as \vspace{-0.2in}
\begin{flalign}
\begin{minipage}{0.4\textwidth}
\begin{eqnarray} \small
\textbf{1} &:& \;
	\begin{pmatrix}
	\phi_1^{\dagger} \phi_1 +\phi_2^{\dagger} \phi_2+\phi_3^{\dagger} \phi_3 
	\end{pmatrix},
\nonumber \\
\textbf{2} &:& \;
	\begin{pmatrix}
	\phi_1^{\dagger} \phi_1 + \omega^2 \phi_2^{\dagger} \phi_2 + \omega \phi_3^{\dagger} \phi_3 \\
	\phi_1^{\dagger} \phi_1 + \omega \phi_2^{\dagger} \phi_2 + \omega^2 \phi_3^{\dagger} \phi_3 
	\end{pmatrix},
\nonumber
\end{eqnarray}
\end{minipage}
\begin{minipage}{0.5\textwidth}
\begin{eqnarray}\small
\textbf{3}: \;
	\begin{pmatrix}
	\phi_1^{\dagger} \phi_2 + \phi_2^{\dagger} \phi_1 \\ 
	\phi_1^{\dagger} \phi_3 + \phi_3^{\dagger} \phi_1 \\ 
	\phi_2^{\dagger} \phi_3 + \phi_3^{\dagger} \phi_2 \\ 
	\end{pmatrix},
\quad
\textbf{3$^{\prime}$}: \;
	\begin{pmatrix}
	-i [\phi_1^{\dagger} \phi_2 -\phi_2^{\dagger} \phi_1] \\
	i [\phi_1^{\dagger} \phi_3 -\phi_3^{\dagger} \phi_1] \\
	-i [\phi_2^{\dagger} \phi_3 -\phi_3^{\dagger} \phi_2] \\ 
	\end{pmatrix}.
	\nonumber 
\end{eqnarray}
\end{minipage}
\nonumber \\
\end{flalign}
Therefore, the $S_4$-invariant 3HDM potential takes on the following form:
\begin{eqnarray}
V_{S_4} &=& -M\,{\bf{ 1}}
	  +\Lambda_{0}\,{\bf{1^{\mathsf{2}}}}
    +\Lambda_{1}\,{\bf{2^{\mathsf{T}}}\cdot \bf{2}}
    +\Lambda_{2}\,{\bf{3^{\mathsf{T}}}\cdot \bf{3}}
    +\Lambda_{3}\,{\bf{3^{\prime\mathsf{T}}}\cdot \bf{3^{\prime}}}.
\end{eqnarray}
Note that the $S_4$-invariant 3HDM potential has a similar form to the $A_4$-invariant 3HDM potential shown in Eq. (\ref{pa4}). However, in the $S_4$-invariant 3HDM potential all parameters are real, due to the absence of ${\bf{3^{\prime \mathsf{T}}}\cdot \bf{3^{}}}$ and ${\bf{3^{\mathsf{T}}}\cdot \bf{3^{\prime}}}$ terms which are allowed by the $A_4$ symmetry. In addition, if Re($\lambda_{1212})=2\lambda_{11}-(\lambda_{1122}+\lambda_{1221})$, one gets an SO(3)-invariant 3HDM potential.

The next symmetry of the 3HDM potential is the $\Delta(54)$ symmetry group, which can be defined through the three generators:
\begin{eqnarray}
\Delta_{\Delta(54)}^1 &=& \text{diag}[{g_1} \otimes \sigma^0,{g_1}^* \otimes \sigma^0], \nonumber \\
\Delta_{\Delta(54)}^2 &=& \sigma^0 \otimes {g_2} \otimes \sigma^0, \nonumber \\
\Delta_{\Delta(54)}^3 &=& \sigma^0 \otimes {g_3} \otimes \sigma^0,
\end{eqnarray}
where
\begin{eqnarray} \small
g_1 = 
	\begin{pmatrix}
	\omega^2& 0 & 0 \\
	0 & \omega & 0\\
	0 & 0 &1
	\end{pmatrix},
\quad
g_2 = 
	\begin{pmatrix}
	0 & 1 & 0 \\
	0 & 0 & 1\\
	1 & 0 & 0
	\end{pmatrix},
\quad
g_3 =
	\begin{pmatrix}
	0 & 1 & 0 \\
	1& 0 &0\\
	0 &0 &1
	\end{pmatrix}.
\end{eqnarray}
These generators satisfy the conditions: $g_1^3=1$, $g_2^3=1$ and $g_3^2=1$. 
The $\textbf{3} \otimes \textbf{3}$ tensor product of $\Delta(54)$ can be decomposed in one singlet \textbf{1} and four doublets \textbf{2}, \textbf{2$^{\prime}$}, \textbf{2$^{\prime\prime}$} \ and \textbf{2$^{\prime\prime\prime}$} as
\begin{eqnarray}
\Delta(54): \quad \textbf{3} \otimes \textbf{3}=\textbf{1} \oplus \textbf{2} \oplus \textbf{2$^{\prime}$} \oplus \textbf{2$^{\prime\prime}$} \oplus \textbf{2$^{\prime\prime\prime}$}.
\end{eqnarray}
The irreducible representations of $\Delta(54)$ in the bilinear-space can be represented as~\cite{GAP}\vspace{-0.2in}
\begin{flalign}
\begin{minipage}{0.4\textwidth}
\begin{eqnarray} \small
\textbf{1} &:& \;
	\begin{pmatrix}
	\phi_1^{\dagger} \phi_1 +\phi_2^{\dagger} \phi_2+\phi_3^{\dagger} \phi_3 
	\end{pmatrix},
\nonumber \\
\textbf{2} &:& \;
	\begin{pmatrix}
	\phi_1^{\dagger} \phi_1 + \omega \phi_2^{\dagger} \phi_2 + \omega^2 \phi_3^{\dagger} \phi_3 \\
	\phi_1^{\dagger} \phi_1 + \omega^2 \phi_2^{\dagger} \phi_2 + \omega \phi_3^{\dagger} \phi_3 
	\end{pmatrix},
\nonumber
\end{eqnarray}
\end{minipage}
\begin{minipage}{0.4\textwidth}
\begin{eqnarray}\small
\textbf{2$^{\prime}$} &:& \;
	\begin{pmatrix}
	\phi_1^{\dagger} \phi_2 + \phi_3^{\dagger} \phi_1 + \phi_2^{\dagger} \phi_3 \\
	\phi_2^{\dagger} \phi_1 + \phi_1^{\dagger} \phi_3 + \phi_3^{\dagger} \phi_2 
	\end{pmatrix},
\nonumber \\
\textbf{2$^{\prime\prime}$} &:& \;
	\begin{pmatrix}
	\phi_2^{\dagger} \phi_3 + \omega \phi_3^{\dagger} \phi_1 + \omega^2 \phi_1^{\dagger} \phi_2 \\
	\omega^2 \phi_2^{\dagger} \phi_1 + \phi_3^{\dagger} \phi_2 + \omega \phi_1^{\dagger} \phi_3 
	\end{pmatrix},
\nonumber \\
\textbf{2$^{\prime\prime\prime}$} &:& \;
	\begin{pmatrix}
	\omega \phi_2^{\dagger} \phi_1 + \phi_3^{\dagger} \phi_2 + \omega^2 \phi_1^{\dagger} \phi_3 \\
	\omega \phi_2^{\dagger} \phi_3 + \omega^2 \phi_3^{\dagger} \phi_1 +  \phi_1^{\dagger} \phi_2 
	\end{pmatrix}.
\nonumber
\end{eqnarray}
\end{minipage}
\nonumber \\
\end{flalign}
Having obtained the $\Delta(54)$-symmetric blocks, the
$\Delta(54)$-invariant 3HDM potential is given by 
\begin{eqnarray}
V_{\Delta(54)} =&& -M{\bf{ 1}}
	  +\Lambda_{0}\,{\bf{1^{\mathsf{2}}}}
    +\Lambda_{1}\,{\bf{2^{\dagger}}\cdot \bf{2}}
    +\Lambda_{2}\,{\bf{2^{\prime\mathsf{T}}}\cdot \bf{2^{\prime}}}
     \nonumber\\
    &&+\Lambda_{3}\,{\bf{2^{\prime\prime\dagger}}\cdot \bf{2^{\prime\prime}}}
    +\Lambda_{4}\,{\bf{2^{\prime\prime\prime\dagger}}\cdot \bf{2^{\prime\prime\prime}}},
\end{eqnarray}
and can be rewritten in the following form:
\begin{align}
V_{\Delta(54)} = &
 - \mu_1^2 \left( |\phi_1|^2\: +\: |\phi_2|^2\: +\: | \phi_3 |^2 \right) 
 + \lambda_{11} \left( |\phi_1|^4 + |\phi_2|^4\: +\: |\phi_3|^4
   \right)\
\nonumber \\ &
+ \lambda_{1122} \left( |\phi_1|^2 |\phi_2|^2 
+ |\phi_1|^2  |\phi_3 |^2 
 + |\phi_2|^2 \,|\phi_3|^2 \right)
\nonumber \\ &
+\lambda_{1221} \left( |\phi_1^{\dagger} \phi_2 |^2
+ |\phi_1^{\dagger} \phi_3|^2
+ |\phi_2^{\dagger} \phi_3|^2\right)
\nonumber \\ &
+ \lambda_{1213} \left( (\phi_1^{\dagger} \phi_2) (\phi_1^{\dagger} \phi_3)
+ (\phi_2^{\dagger} \phi_1) (\phi_2^{\dagger} \phi_3) 
+ (\phi_3^{\dagger} \phi_2) (\phi_3^{\dagger} \phi_1) \right)
\nonumber \\ &
+ \lambda^*_{1213} \left( (\phi_2^{\dagger} \phi_1) (\phi_3^{\dagger} \phi_1)
+ (\phi_1^{\dagger} \phi_2) (\phi_3^{\dagger} \phi_2) 
+ (\phi_2^{\dagger} \phi_3) (\phi_1^{\dagger} \phi_3)  \right).
\end{align}

The largest non-Abelian discrete symmetry of 3HDM is the $\Sigma(36)$
group
\cite{Ivanov:2012fp,Keus:2013hya,Ivanov:2014doa,deMedeirosVarzielas:2019rrp}. The
relevant $\textbf{3} \otimes \textbf{3}$ tensor product of
$\Sigma(36)$ may be decomposed as 
\begin{eqnarray}
\Sigma(36): \quad \textbf{3} \otimes \textbf{3} = \textbf{1} \oplus \textbf{4} \oplus \textbf{4$^{\prime}$}.
\end{eqnarray}
The three generators of this group are
\begin{eqnarray}
\Delta_{\Sigma(36)}^1 &=& \text{diag}[{g_1} \otimes \sigma^0,{g_1}^* \otimes \sigma^0], \nonumber \\
\Delta_{\Sigma(36)}^2 &=&\sigma^0 \otimes {g_2} \otimes \sigma^0, \nonumber \\
\Delta_{\Sigma(36)}^3 &=& \text{diag}[{g_3} \otimes \sigma^0,{g_3}^* \otimes \sigma^0], 
\end{eqnarray}
with
\begin{eqnarray} \small
g_1 = 
	\begin{pmatrix}
	1 & 0 & 0 \\
	0 & \omega & 0 \\
	0 & 0 & \omega^2
	\end{pmatrix},
\quad
g_2 = 
	\begin{pmatrix}
	0 & 1 & 0 \\
	0 & 0 & 1 \\
	1 & 0 & 0
	\end{pmatrix},
\quad
g_3 = {1 \over \omega-\omega^2}
	\begin{pmatrix}
	1 & 1 & 1 \\
	1 & \omega & \omega^2 \\
	1 & \omega^2 & \omega
	\end{pmatrix},
\end{eqnarray}
which obey the conditions: $g_1^3=g_2^3=1$ and $g_3^4=1$. Now, we can
define the irreducible representations of $\Sigma(36)$ as~\cite{GAP}
$$ \small
\textbf{1}: \;
	\begin{pmatrix}
	\phi_1^{\dagger} \phi_1 +\phi_2^{\dagger} \phi_2+\phi_3^{\dagger} \phi_3 
	\end{pmatrix},
$$
\begin{flalign}
\begin{minipage}{0.4\textwidth}
\begin{eqnarray}\small
\textbf{4} &:& \;
	\begin{pmatrix}
	\omega^2 \phi_1^{\dagger} \phi_2 + \phi_2^{\dagger} \phi_3 + \omega \phi_3^{\dagger} \phi_1 \\
	\omega^2 \phi_1^{\dagger} \phi_3 + \phi_2^{\dagger} \phi_1 + \omega \phi_3^{\dagger} \phi_2 \\
	\omega^2 \phi_1^{\dagger} \phi_2 + \omega \phi_2^{\dagger} \phi_3 + \phi_3^{\dagger} \phi_1 \\
	\omega^2 \phi_1^{\dagger} \phi_3 + \omega \phi_2^{\dagger} \phi_1 + \phi_3^{\dagger} \phi_2 
	\end{pmatrix},
\nonumber
\end{eqnarray}
\end{minipage}
\begin{minipage}{0.4\textwidth}
\begin{eqnarray}\small
\textbf{4$^{\prime}$} &:& \;
	\begin{pmatrix}
	\omega \phi_1^{\dagger} \phi_3 + \omega \phi_2^{\dagger} \phi_1 + \omega \phi_3^{\dagger} \phi_2 \\
	\omega \phi_1^{\dagger} \phi_1 + \phi_2^{\dagger} \phi_2 + \omega^2 \phi_3^{\dagger} \phi_3 \\
	\omega^2 \phi_1^{\dagger} \phi_2 + \omega^2 \phi_2^{\dagger} \phi_3 + \omega^2 \phi_3^{\dagger} \phi_1 \\
	\omega \phi_1^{\dagger} \phi_1 + \omega^2 \phi_2^{\dagger} \phi_2 + \phi_3^{\dagger} \phi_3 
	\end{pmatrix}.
\nonumber 
\end{eqnarray}
\end{minipage}
\nonumber \\
\end{flalign}
Finally, the $\Sigma(36)$-invariant potential then takes on the form
\begin{eqnarray}
V_{\Sigma(36)} &=& -M\,{\bf{ 1}}
	  +\Lambda_{0}\,{\bm{1}}^{\mathsf{2}}
    +\Lambda_{1}\,{\bf{4^{\dagger}}\cdot \bf{4}}
    +\Lambda_{2}\,{\bf{4^{\prime\dagger}}\cdot \bf{4^{\prime}}}.
\end{eqnarray}
Equivalently, the $\Sigma(36)$-invariant 3HDM potential can be given by
\begin{align}
V_{\Sigma(36)} = &
- \mu_1^2 \left( |\phi_1|^2\: +\: |\phi_2|^2\: +\: | \phi_3 |^2 \right) 
 + \lambda_{11} \left( |\phi_1|^4 + |\phi_2|^4\: +\: |\phi_3|^4
   \right)\
\nonumber \\ &
+ \lambda_{1122} \left( |\phi_1|^2 |\phi_2|^2 
+ |\phi_1|^2  |\phi_3 |^2 
 + |\phi_2|^2 \,|\phi_3|^2 \right)
\nonumber \\ &
+ \lambda_{1221} \left( |\phi_1^{\dagger} \phi_2 |^2
+ |\phi_1^{\dagger} \phi_3|^2
+ |\phi_2^{\dagger} \phi_3|^2\right)
\nonumber \\ &
+{3\over 4}(2\lambda_{11}-\lambda_{1122}-\lambda_{1221})\nonumber \\ & \times \left( (\phi_1^{\dagger} \phi_2) (\phi_1^{\dagger} \phi_3)
+ (\phi_1^{\dagger} \phi_3) (\phi_2^{\dagger} \phi_3) 
+ (\phi_1^{\dagger} \phi_2) (\phi_3^{\dagger} \phi_2) + {\rm H.c.}\right),
\end{align}
with all real parameters.

These symmetries for the 3HDM potential, along with their non-zero
theoretical parameters, are presented in Table~\ref{tab2}. 

\newpage

\end{document}